\documentclass[]{aa}
\usepackage{graphicx,amsmath}
\usepackage{txfonts}

\newcommand{\st}{{\it St}}

\begin{document}
\title{On the  growth and evolution of low-mass planets in pressure bumps}
\titlerunning{Evolution of planets in pressure bumps}
\author{Arnaud Pierens
 \inst{1}
 \and
  Sean N. Raymond \inst{1}}
\institute{ Laboratoire d'astrophysique de Bordeaux, Univ. Bordeaux, CNRS, B18N, all\'ee Geoffroy Saint-Hilaire, 33615 Pessac, France\\
\email{arnaud.pierens@u-bordeaux.fr}
}

\abstract{
Observations of protoplanetary discs have revealed dust rings which are likely due to the presence of pressure bumps in the disc.  Because these structures tend to trap drifting pebbles, it has been proposed that pressure bumps may play an important role in the planet formation process. In this paper, we investigate the  orbital evolution of a $0.1$ $M_\oplus$ protoplanet embedded in a pressure bump using 2-dimensional hydrodynamical simulations of protoplanetary discs consisting of gas and pebbles. We  examine  the role of thermal forces generated by the pebble accretion-induced heat release, taking into account the feedback between luminosity and eccentricity. We also study the effect of the pebble-scattered flow on the planet's orbital evolution. 
 Due to accumulation of pebbles  at the  pressure bump,  the planet's accretion luminosity is  high enough to induce significant eccentricity growth through thermal forces. Accretion luminosity is also responsible for vortex formation at the planet position through baroclinic effects,  which cause the planet escape from the dust ring if dust feedback onto the gas is neglected.  Including the effect of the dust back-reaction leads to weaker vortices, which enable the planet to remain close to the pressure maximum on an eccentric orbit. Simulations in which the planet mass is allowed to increase as a consequence of pebble accretion resulted in the formation of giant planet cores with mass in the range $5-20$  $M_\oplus$ over $\sim 2\times 10^4$ yrs. This occurs   for moderate values of the Stokes number $\st \approx 0.01$ such that the pebble drift velocity is not too high and the dust ring mass not too small. Our results suggest that pressure bumps  mays be preferred locations for the formation of giant planets, but this requires a moderate level of grain growth within the disc. 
 }


\maketitle

\section{Introduction}

In the standard scenario for planet formation, mm-sized dust grows  from the coagulation of micron-sized grains (Dullemond \& Dominik 2005).  However, further growth  of particles in the mm-size range is limited because of the bouncing  (Zsom et al. 2010) and fragmentation (Blum \& Wurm 2008) barriers.   An emerging picture to bypass these growth barriers is that 100-km sized planetesimals form directly through the streaming instability (Youdin \& Goodman 2005; Johansen et al. 2009; Simon et al. 2016) as particles with Stokes number (or dimensionless stopping time) $\st\sim 0.001-0.1$  directly concentrate into clumps or filaments under the action of gas drag, and these  become gravitationally unstable  to form $\sim100-1000$ km size bodies. Planetesimals can subsequently grow very efficiently by capturing inward drifting pebbles (Johansen \& Lacerda 2010; Lambrechts \& Johansen 2012), namely solids with Stokes number $\st\sim 0.01-1$ that are marginally coupled to the gas,  leading eventually to the formation of  giant planet cores within 1 Myr (Lambrechts \& Johansen 2014). \\

The conditions for triggering the SI, and the efficiency of pebble accretion, are however sensitive to the level of turbulence operating in the disc. Small particles are lofted away from the midplane by turbulent mixing, such that the local dust-to-gas ratio, and hence the growth rate of the SI, are reduced. One possibility to counteract the effect of turbulence in the disc is to invoke the presence of pressure bumps in the disc, where the radial drift of dust particles can be stopped, leading subsequently to an enhancement of the dust-to-gas ratio.

Pressure bumps may be associated with ringed structures,  which have been observed with ALMA in a variety of protoplanetary discs (ALMA Partnership et al. 2015; Andrews et al. 2016,2018). Although there is no current consensus, possible mechanisms that might be able to explain these features include: the presence of planets that can open gaps in the disc (Dipierro et al. 2015; Dong et al. 2017), zonal flows (Flock et al. 2015), snow lines (Zhang et al. 2015), large-scale instabilities (L\`oren-Aguilar \& Bate 2016), 
spontaneous ring formation through radial drift plus dust coagulation (Gonzalez et al. 2017), and dust-driven viscous ring-instability (Dullemond \& Penzlin 2018). Whenever the physical origin of pressure bumps, recent work suggests that planetesimals could indeed be formed at such locations as a result of the SI (Carrera et al. 2021; Xu \& Bai 2022), at least if the protoplanetary disc can form grains larger than 1mm (Carrera et al. 2022). 

 Planetesimals formed at pressure bumps are expected to grow efficiently though pebble accretion.  It has been shown that the gravitational  potential energy release associated with pebble accretion  can significantly impact its orbital evolution in the disc (Benitez-Llambay et al. 2015). Accretion heating  makes gas streamlines outflowing from the Hill sphere form  two underdense lobes -- one leading and one following -- the planet that exert a torque on the planet. Because the orbital velocity of gas is slightly sub-Keplerian, the lobe located behind the planet tends to be less dense than the one located ahead of the planet, resulting in a net positive torque, which has been referred to as the "heating torque". Due to the locally enhanced dust-to-gas ratio at the pressure bump, the heating torque could  potentially be strong, as well as the torque induced by scattered pebble flow (Benitez-Llambay \& Pessah 2018). It can not be excluded that  a planet's orbital evolution would even be  dominated by these processes, since the classical Lindblad and corotation torques arising from the interaction with the gas disc tend to counteract each other in the vicinity of the pressure bump (Masset et al. 2006).  Of course, this would be true provided that the equilibrium planet position set by the Lindblad and corotation torques lies within the pebble ring, otherwise the pebble accretion rate onto the planet would be very small (Morbidelli 2020).  The effect of thermal torques on the migration of Mars- to super-Earth-sized planets in the vicinity of a pressure bump was recently investigated by Chrenko \& Chametla (2023). These authors found that thermal forces can cause low-mass planets to escape the pressure bump. This arises because thermal forces make the planet eccentricity grow, which subsequently quenches the corotation torque responsible for the migration trap. 
 
In this paper, we  use 2D hydrodynamical simulations to examine the orbital evolution of a protoplanet  forming at a pressure bump and whose mass increases as a result of pebble accretion. Torques induced by the pebble scattered flow are taken into account and the pebble accretion rate is determined self-consistently using the formulae of Liu \& Ormel (2018) and Ormel \& Liu (2018) for the pebble accretion efficiency.  This is an important issue as the latter depends on the planet eccentricity, while the planet eccentricity growth rate through thermal forces is regulated by the accretion rate. Hence, the positive feedback that exists between the planet luminosity and eccentricity (Velasco Romero et al. 2022) is also considered on our study. The aim of this paper is twofold:  first, we identify the main physical processes that drive the planet orbital evolution at a pressure bump. Second, we study whether giant planet cores can  form at a pressure bump under realistic conditions for the accretion and migration of the planet.

This paper is organized as follows. In Sect. 2 and 3, we present the physical model and the numerical setup.  In Sect. 4.,  we  discuss the results of our simulations.  Finally,  we draw our conclusions in Sect. 5. 

\section{The physical model}
\label{sec:dustdisc}

  \subsection{Gas equations}
We solve the hydrodynamical equations for the gas component in polar coordinates $(r,\varphi)$ (radial, azimuthal), with the origin of the frame located at the central star.   The governing equations  are the continuity equation which is given by:

\begin{equation}
\frac{\partial \Sigma_g}{\partial t}+\nabla\cdot(\Sigma_g{\mathbf v_g})=0,
\end{equation}

the Navier-Stokes equation which is given by:
\begin{equation}
\frac{\partial {\mathbf v_g}}{\partial t}+({\mathbf v_g}\cdot \nabla){\mathbf v_g}=-\frac{\nabla P}{\Sigma}-\frac{\Sigma_d}{\Sigma_g}\mathbf F_{drag}-{\bf \nabla} \Phi-\frac{\nabla\cdot \cal T}{\Sigma_g},
\label{eq:gas}
\end{equation}

and the energy equation which is described in Sect. \ref{sec:energy}. In Eq. \ref{eq:gas},  $\Sigma_g$ and $\Sigma_d$ are the gas and pebble surface densities respectively, $\mathbf v_g$ the gas velocity, $\mathbf F_{drag}$ describes the frictional drag force  between the gas and the dust (see Sect.  \ref{sec:dustequations}), $P$ the gas pressure and $\cal T$ the viscous stress tensor (e.g. Nelson et al. 2000).

\subsubsection{Gravitational potential}

In Eq. \ref{eq:gas}, the gravitational potential $\Phi$  
includes the contributions from the star, planet and the indirect term. In this work, the gravitational influence of the planet on the gas disc is modelled using  a vertically averaged expression for the gravitational potential $\Phi_p$ (Muller \& Kley 2012, Chrenko et al. 2017): \\

\begin{equation}
\Phi_p=\frac{-Gm_p}{\Sigma_g}\int_{-z_{max}}^{z_{max}} \frac { \rho_g(z) dz }{\sqrt{|{\bf r}-{\bf r_p}|^2+r_s^2}}+\frac{Gm_p}{r_p^3}{\bf r}\cdot{\bf r_p}
\label{eq:phi_p}
\end{equation}

where $m_p$ is the planet mass,  $r_s$ is a softening length which is employed to avoid divergence of the gravitational potential at the planet location,  ${\bf r }$ is the vector pointing to the location in the disk  and  ${\bf r_p }$ is the vector pointing to the planet.   In the previous equation, the second term corresponds to the acceleration experienced by the centre of the reference frame due to the presence of the planet. The integral in Eq. \ref{eq:phi_p}  is computed by dividing the interval $[0,z_{max}]$ into $N_z=10$ equal intervals with $z_{max}=3H$ where $H$ is the disc scale height,  and assuming hydrostatic equilibrium in the vertical direction such that the mass density of the gas $\rho_g(z)$ is given by:
\begin{equation}
\rho_g(z)=\frac{\Sigma_g}{\sqrt{2\pi}H}\exp\left(-\frac{z^2}{2H^2}\right)
\end{equation}

 This averaging procedure allows us to use a smoothing length $r_s$ similar to  the characteristic grid cell at $r_p$, and which is set to $r_s=0.01$ $H$ in this work. This is much smaller than the value for the softening parameter that is classically adopted in 2d simulations,  which is chosen to be of the order of $H$ to mimic the gravitational potential of a vertically extended 3D disc. It has been shown, however, that leads to torques experienced by the planet that can significantly differ from the full 3d case (Muller \& Kley 2012). 

The indirect term $\Phi_{ind}$ arising from the fact that the frame centred on the central star is not inertial  is given by (e.g. Nelson et al. 2000):

\begin{equation}
\Phi_{ind}=G\int_S\frac{(\Sigma_g+\Sigma_d)d{S'}}{R'^3}{\bf R}\cdot{\bf R'}
\end{equation}

where $S$ is the surface of the simulation domain.

\subsubsection{Energy equation}
\label{sec:energy}

In addition to the continuity equation and momentum equation, we solve an energy equation that includes the effect of 
viscous heating, stellar irradiation, and radiative cooling. It reads:
\begin{equation}
\frac{\partial e}{\partial t}+\nabla \cdot (e{\bf v_g})=-(\gamma-1)e{\nabla \cdot {\bf v_g}}+Q^+_{vis}-Q^--2H\nabla  \cdot {\bf F}+Q_{acc}
\label{eq:energy}
\end{equation}
where $e$ is the thermal energy density,  $\gamma$ the adiabatic index which is set 
to $\gamma=1.4$. In the previous equation,  $Q^+_{vis}$ is the viscous heating term,  and $Q^-=2\sigma_B T_{eff}^4$ is the local radiative cooling from the disc surfaces, where 
$\sigma_B$ is the Stephan-Boltzmann constant and $T_{eff}$ the effective temperature which is given by (Menou \& Goodman 2004):
\begin{equation}
T_{eff}^4=\frac{T^4-T_{irr}^4}{\tau_{eff}} \quad \text{with} \quad \tau_{eff}=\frac{3}{8}\tau+\frac{\sqrt{3}}{4}+\frac{1}{4\tau}
\end{equation}
Here, $T$ is the midplane temperature and $\tau=\kappa\Sigma_g/2$ is the vertical optical depth, where  $\kappa$ is the Rosseland mean opacity which is taken from Bell \& Lin (1994). $T_{irr}$ is the irradiation temperature which is computed from the irradiation flux (Menou \& Goodman 2004):
\begin{equation}
\sigma_B T_{irr}^4= \frac{L_\star(1-\epsilon)}{4\pi R^2}\frac{H}{R}\left(\frac{d \log H}{d\log R}-1\right)
\end{equation}
where $\epsilon=1/2$ is the disc albedo, $L_\star$  is the stellar luminosity which is set to $L_\star=1.43 L_\odot$ , and where the factor $d \log H/d\log R$ is set to 
be $d \log H/d\log R=9/7$ (Chiang \& Goldreich 1997). This implies that self-shadowing effects are not taken into account in this study.\\
In Eq. \ref{eq:energy}, ${\bf F}$ is the radiative flux which is treated in the flux-limited  diffusion approach and which 
reads (e.g. Kley \& Crida 2008):
\begin{equation}
{\bf F}=-\frac{16\sigma_B \lambda T^3}{\rho_g \kappa}\nabla T
\end{equation}
 where $\lambda$ is a flux-limiter (e.g.  Kley 1989). Finally, $Q_{acc}=L_{acc}/S$ is the accretion heating term, with $S$ the cell area and $L_{acc}$ the luminosity of the accreting embryo (see Sect. \ref{sec:planets}).

  \subsection{Dust equations}
  \label{sec:dustequations}

The solid component is treated as a pressureless fluid whose equations for the conservation of mass and momentum are given by:

\begin{equation}
\frac{\partial \Sigma_d}{\partial t}+\nabla\cdot(\Sigma_d{\mathbf v_d})=-\dot \Sigma_d
\label{eq:dust_continuity}
\end{equation}

and

\begin{equation}
\frac{\partial {\mathbf v_d}}{\partial t}+({\mathbf v_d}\cdot \nabla){\mathbf v_d}=\mathbf F_{drag}-{\bf \nabla} \Phi -\frac{D_v}{t_s}\frac{\nabla \Sigma_d}{\Sigma_d}
\label{eq:dust}
\end{equation}

where $\mathbf v_d$ is the dust velocity, $t_s$ is the particle stopping time, $D_v$ is the turbulent (viscous) dust diffusion coefficient, and $\dot \Sigma_d$ represents the local dust density decrease resulting from pebble accretion (see Sec. \ref{sec:planets}).
 \subsubsection{Drag force between gas and dust}

In Eq. \ref{eq:dust}, the drag force resulting from the interaction with the gaseous disc is given by:
\begin{equation}
\mathbf F_{drag}=\frac{1}{t_s}({\mathbf v_g -\mathbf v_p})
\end{equation}
In the following, we parametrize the stopping time $t_s$  through the Stokes number $\st =t_s \Omega$ with $\Omega$ the Keplerian frequency. We  note that in the well-coupled regime, $t_s$ is related to the particle size $s_d$  through:
\begin{equation}
t_s=\frac{s_d\rho_d}{\Sigma_g \Omega}
\end{equation}

where $\rho_d$ is the particle internal density. 
With respect to our previous work where we employed a model of dust evolution with a single representative, we here rather adopt a fixed particle Stokes number, for which we considered values of $\st=0.001, 0.01, 0.05$.

 \subsubsection{Dust diffusion}

 Tominaga et al. (2019) showed that introducing a diffusion term as a source term in the continuity equation for the dust can violate the conservation of the total angular momentum of the dusty  disc. To bypass this issue,  Klahr \& Schreiber (2020)  implemented the diffusion flux in the momentum equation for the dust rather than in the continuity equation. As a consequence, we model dust diffusion arising from the residual turbulence in the disc by including a diffusion pressure term within the dust momentum equation and corresponding to the last source term in Eq. \ref{eq:dust}. The corresponding dust diffusion coefficient is given by:

\begin{equation}
D_v=\frac{1+\st+4\st^2}{(1+\st^2)^2} \alpha c_s H
\label{eq:d}
\end{equation}

with $\alpha$ the viscous stress parameter (Shakura \& Sunyaev 1973)

 \subsection{Planets}
 \label{sec:planets}
  \subsubsection{Accretion luminosity}
We consider luminous planets that can release heat in the disc as a consequence of  pebble accretion.  The   planet luminosity is related to the pebble accretion rate  onto the planet $\dot M_p$ through the relation:

\begin{equation}
L=\frac{Gm_p\dot M_p }{R_p}
\end{equation}

where $R_p$ is the physical radius of the planet which is calculated assuming a material density of $2$ $g.cm^{-3}$. Defining the pebble accretion efficiency as $\epsilon=\dot M_p/\dot M_{\rm peb}$ where  $\dot M_{\rm peb}$ is the inward mass flux of pebbles, it has been shown that $\epsilon$ depends mainly on $\eta$, $\st$, $\alpha$ and  on the planet eccentricity $e_p$.   Liu \& Ormel (2018) and Ormel \& Liu (2018) have derived prescriptions to calculate $\epsilon$ as a function of these parameters (see Appendix A) and that we employ to self-consistently determine the value of the planet luminosity. Thermal forces arising from heat release in the disc are indeed expected to make the planet eccentricity increase, and consequently impact the pebble accretion efficiency. As noted by Velasco Romero et al. (2021), there exists a feedback loop between the eccentricity growth driven by thermal forces and the luminosity  which  tends to increase with eccentricity, at least in the regime where the eccentricity remains smaller than the disc aspect ratio. \\

  \subsubsection{Accretion procedure}
  \label{sec:accretion}
In our simulations, the pebble accretion rate onto the planet is therefore based on the prescription of Liu \& Ormel (2018) and is given by $\dot M_p=\epsilon \dot M_{\rm peb}$. To make sure that the latter is indeed smaller  to what the disc can provide, we follow Crida \& Bitsch (2017) and calculate the maximum expected accretion rate $\dot M_K$ using Kley (1999)'s recipe.  At each timestep,  the total  mass of pebbles located within 0.5 of the accretion radius $R_{acc}$  of the planet is   added to the planet mass, at such a rate that 1/2 of the total pebble mass within 0.5 $R_{acc}$ is accreted by the planet over an orbital period.   The effective pebble accretion rate onto the planet is then given by: 
\begin{equation}
\dot M_p=\min(\epsilon \dot M_{\rm peb}, \dot M_K)
\end{equation}
Accretion of solids is accompanied by a reduction of the local gas surface density within the accretion radius of the planet  according to the determined value for the accretion rate,  and which is represented  by the source term in the dust continuity equation (see Eq. \ref{eq:dust_continuity}).  We also note that the angular momentum contained within the  removed material is not added to the planet's angular momentum. 
\section{Numerical simulations}
\subsection{Numerical method}
Simulations were performed using the GENESIS (De Val-Borro et al. 2006) code which solves
the equations governing the disc evolution on a polar grid $(R,\varphi)$ using an advection scheme based on the monotonic  transport algorithm (Van Leer 1977). It uses the FARGO algorithm (Masset 2000) to avoid time step limitation due to the Keplerian velocity at the inner edge of the disc, and was recently extended to follow the evolution of a solid component that is modelled assuming a pressureless fluid.  Momentum exchange between the particles and the gas is handled by employing the semi-analytical scheme presented in Stoyanovskaya et al. (2018). This approach enables considering arbitrary solid concentrations and values for the Stokes number, and  is therefore very well suited for looking for solutions of non-stationary problems.  Tests of the numerical method to handle the momentum transfer between gas and dust were presented in 
Pierens et al. (2019) and the code was recently used to study the non-linear evolution of the Secular Gravitational Instability (Pierens 2021).\\

The computational domain is covered by $N_R=1572$ radial grid cells uniformly distributed  between $R_{\rm in}=0.55$  and 
$R_{\rm out}=1.6$, and $N_\phi=3072$ azimuthal grid cells.   The numerical resolution is chosen so that i) the amplitude of the dust torques is converged (see Sect. \ref{sec:dusttorques}) and ii) the thermal disturbance in the vicinity of the planet is resolved by $\sim 5$ grid cells (see Sect. \ref{sec:pebble_accretion}). The computational units that we adopt are such that the unit of mass is the central mass $M_\star=1$ and is assumed to be equivalent to one Solar mass, the gravitational constant is $G=1$, and the distance  $R=1$ in the computational domain is set to  $5.2$ AU. In the following, this corresponds also to the initial semimajor axis $R_0$ of migrating planets so that when presenting the simulation results, time will be expressed in orbital periods at $R=R_0$.
 
 \subsection{Initial and boundary conditions}
 \label{sec:init}

The initial disc aspect ratio $h$ is set assuming that  stellar heating dominates over viscous heating and corresponds to that expected for a purely passive disc 
be $h\propto (R/R_0)^{2/7}$ (Chiang \& Goldreich 1997). For a  constant constant mass flow through the disc $\dot M$, this implies an initial gas surface density   $\Sigma_{BG}=\Sigma_0 (R/R_0)^{-15/14}$ (Bitsch et al. 2014) where $\Sigma_0=6\times 10^{-4}$ is the surface density at $R=R_0$. Overimposed on this background surface density, we set up a Gaussian bump by employing the following profile for the $\alpha$ viscous stress parameter. 
\begin{equation}
\alpha(r)=\frac{\alpha_{BG}}{1+a \exp\left(-\frac{(R-R_0)^2}{2w^2}\right) }
\end{equation}
where $\alpha_{BG}$ is the background value for which we consider values in the range  $[10^{-5}, 10^{-3}]$ and $a=0.5$ is the amplitude of the pressure bump. Hence, the initial surface density profile is given: 
\begin{equation}
\Sigma_g(R)=\Sigma_{BG} \left[ 1+a \exp\left(-\frac{(R-R_0)^2}{2w^2}\right)\right]
\end{equation}

The dust surface density was initialized  assuming a constant dust-to-gas ratio $Z=0.01$. 

Our simulations are split into two steps. We first evolve the gas and pebble discs until a quasi-stationary state is achieved, namely until the gas  and solid mass fluxes become constant through the disc, and then use the relaxed hydrodynamical quantities as initial conditions for runs with embedded planets.  In that case, we make use of wave-killing zones in the intervals $R\in [0.55,0.6]$ and $R\in [1.55,1.6]$ where each  variable f  is relaxed towards its initial value $f_0$ using the prescription of  de Val-Borro et al. (2006):

\begin{equation}
\frac{df}{dt}=-\frac{f-f_0}{\tau}{\cal P}(R)
\end{equation}

where   $\tau$ is the damping time-scale which is set to $0.1$ of the orbital period, and ${\cal P}(R)$ is a parabolic function which varies  between $1$ and $0$ from the edge of the domain to the inner edge of the damping zone.

\section{Results}
\subsection{A reference run}
\label{sec:fiducial}

\begin{figure}
\centering
\includegraphics[width=\columnwidth]{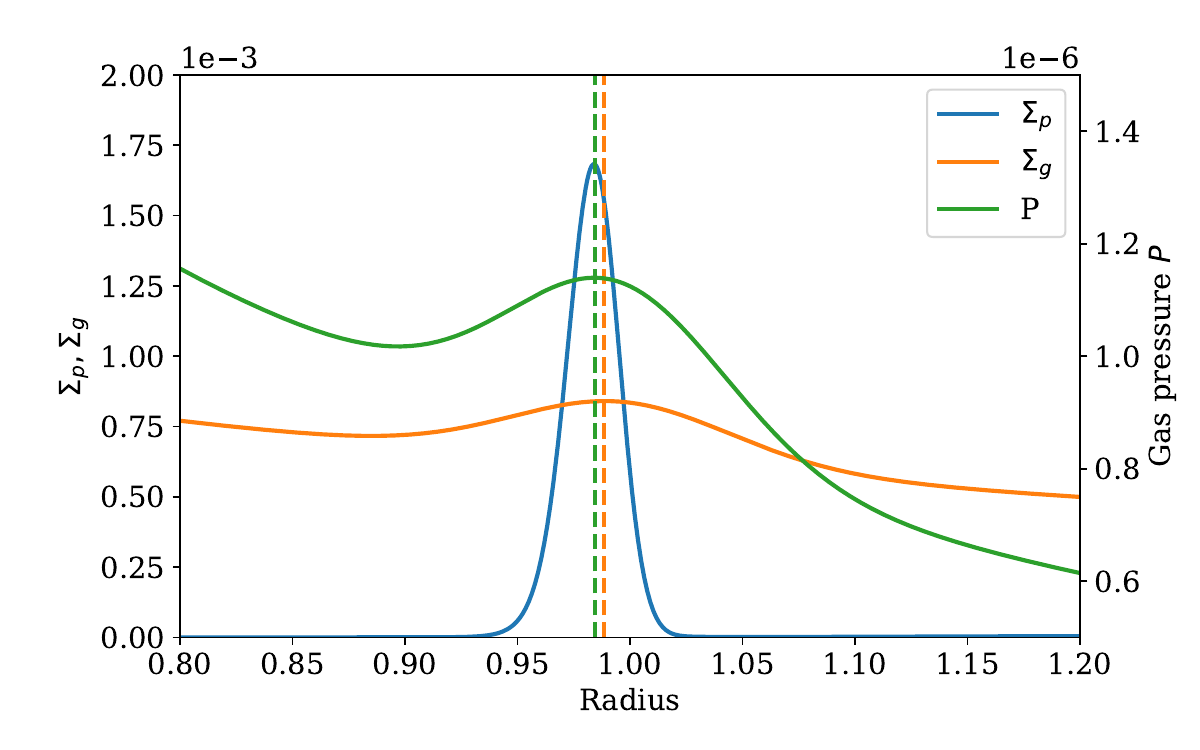}
\caption{Pebble surface density $\Sigma_p$, gas surface density $\Sigma_g$, and gas pressure as a function of radius for the fiducial run with $\st=0.01$ and $\alpha=10^{-4}$.  }
\label{fig:morby}
\end{figure}

\begin{figure}
\centering
\includegraphics[width=\columnwidth]{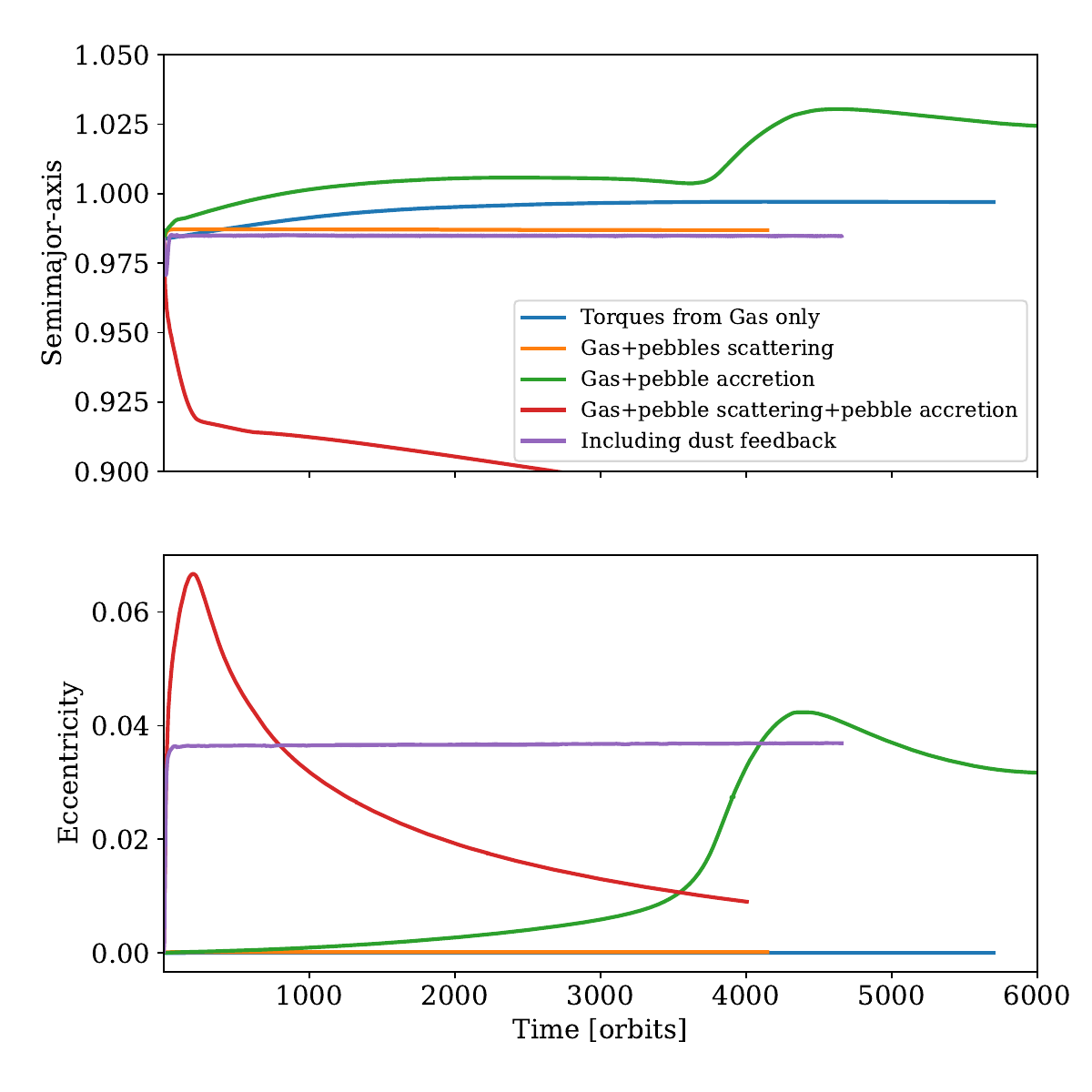}
\caption{Planet semi-major axis and eccentricity as a function of time for the fiducial run with $\st=0.01$ and $\alpha=10^{-4}$ and in cases where a $0.1$ $M_\oplus$ protoplanet is allowed to evolve under the effect of gas torques, pebble scattering, pebble accretion, and dust feedback.}
\label{fig:base}
\end{figure}

\begin{figure}
\centering
\includegraphics[width=\columnwidth]{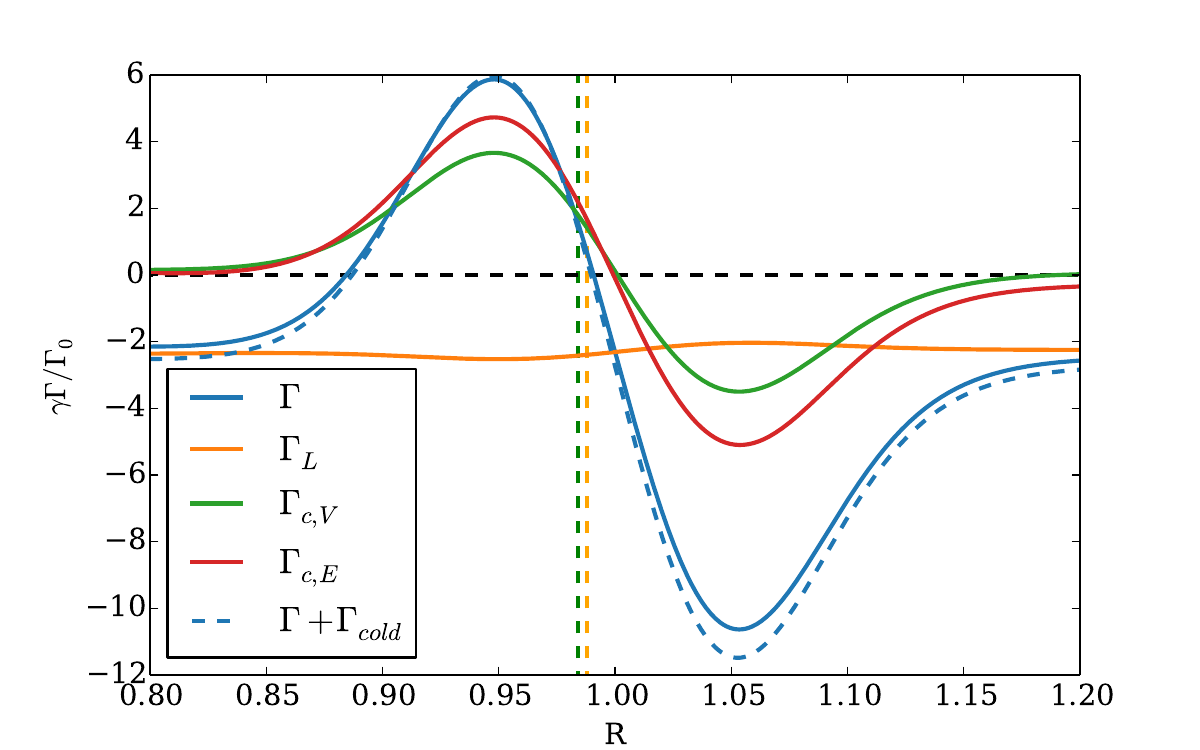}
\caption{Torque profile for the fiducial simulation with $\st=0.01$ and $\alpha=10^{-4}$ at a point in time when migration is halted. Overplotted are the contributions from the Lindblad torque $\Gamma_L$, vortensity-related corotation torque $\Gamma_{c,V}$, and entropy-related corotation torque $\Gamma_{c,V}$. The dashed line shows the total torque in the case where the cold torque $\Gamma_{cold}$ is taken into account.}
\label{fig:torques}
\end{figure}

In the following, we  focus on a fiducial model which has $\st=0.01$ and $\alpha=10^{-4}$. The model is evolved for $5\times 10^5$ orbits  prior to inserting the planet, which is typically the time required to reach a quasi-stationary configuration.  The  dust and gas surface densities  and pressure radial profiles at equilibrium are shown in  Fig. \ref{fig:morby}. As expected, dust trapping at the pressure bump leads to the formation of a ring whose mass is  $\sim 15$ $M_\oplus$. We also see that the pressure and gas surface density maxima are slightly shifted because of the use of a non constant sound speed. The pressure maximum is indeed located at $R\approx 0.984$ whereas the surface density maximum is found to be at $R\approx 0.988$.

 At Time=$5\times 10^5$, we  insert a $0.1$ $M_\oplus$ core  that  initially evolves on a nearly circular orbit  with initial semi-major axis $a_p=1$, eccentricity  $e_p=10^{-4}$, and we examine its orbital evolution within the pressure bump. In order to highlight the physical processes that control the dynamics of an embryo formed  at a pressure bump, we adopted a step-by-step procedure. We first  performed a simulation where we assumed that the protoplanet  feels only the effects of the  gas torques.  Next, we studied the impact of taking into account the   torques resulting from the scattered pebble flow (Benitez-Llambay \& Pessah 2018). Then, we performed an additional run with accretion heating included. Finally, we performed  full calculations in which each of the aforementioned processes are taken into account.  The results of this multistep procedure are shown in Fig. \ref{fig:base}, where we plot for each run the planet  semimajor axis and eccentricity as a function of time. 

\subsubsection{Orbital evolution under the effect of gas torques only}

 We first focus on the situation where only the effects of the gas torques are considered.  The total gas torque $\Gamma$ exerted on the planet is given by: 
 \begin{equation}
 \Gamma=\Gamma_L+\Gamma_{c,V}+\Gamma_{c,E}
 \end{equation}
 
 where $\Gamma_L$, $\Gamma_{c,V}$, $\Gamma_{c,E}$ are the Lindblad torque, vortensity-driven corotation torque, and entropy-related corotation torque respectively. Following Morbidelli (2020), these components can be written as: 
 
 \begin{equation}
 \Gamma_L=(-2.5-1.7b+0.1a) \frac{\Gamma_0}{\gamma},
 \end{equation}
 
  \begin{equation}
 \Gamma_{c,V}=1.1 f_V(\nu,q)\left(\frac{3}{2}-a\right) \frac{\Gamma_0}{\gamma},
 \end{equation}
 
 and 
 
  \begin{equation}
 \Gamma_{c,E}= f_E(\nu,\chi, q)\left(7.9\frac{\xi}{\gamma}\right) \frac{\Gamma_0}{\gamma},
 \end{equation}
 
 where $a$, $b$ are the power-law indexes of the surface density and temperature profiles respectively,  and $\xi=b-(\gamma-1)a$ the  exponent of the entropy profile. $f_V$ and $f_E$ are saturation functions that depend on planet mass ratio $q=m_p/M_\star$, disc viscosity $\nu$ and thermal diffusivity $\chi$ and can be computed using the classical Paardekooper et al. (2011) formulae. In the previous equations, $\Gamma_0$ is the nominal torque which is given by:
 
  \begin{equation}
 \Gamma_{0}=q^2h^2\Sigma_g a_p^2 \Omega_p^2
 \end{equation}
 
 with $\Omega_p$ the planet angular velocity. When only the contribution from the gas torques is taken into account, Fig. \ref{fig:base} shows that the planet becomes trapped near the surface density maximum  at a radial location $r_p\approx 0.995$. Following Morbidelli (2020) and defining the location of the trap relative to the pressure bump as $\delta \omega^2/r_0$, this would correspond to $\delta \approx 6$. It is worth noticing that the planet trap  is located slightly beyond the surface density  maximum, in a region where the surface density gradient is $<0$.  Although the vortensity-related corotation torque is not expected  to be very strong at this location,  the (total) corotation torque  may still counterbalance the Lindblad torque provided that the contribution from  the entropy-related corotation torque is $>0$ and unsaturated near the pressure bump. To demonstrate that this is indeed the case,  $\Gamma_L$, $\Gamma_{c,V}$ and $\Gamma_{c,E}$ are represented as a function of radius in Fig. \ref{fig:torques}.   Consistently with the results of our simulations, we see that the total torque is expected to cancel outside of the maximum gas density. From Fig. \ref{fig:torques}, this is clearly a consequence of the strong entropy-related corotation torque which is  indeed $>0$ in the vicinity of the pressure bump and which acts in concert with $\Gamma_{c,V}$ to counteract the effect of the Lindblad torque.  In the case where the entropy-related corotation torque was saturated, however, we would expect the trap to be located in a  region of positive surface density gradient, since a strong vortensity-related corotation torque would be required to cancel the Lindblad torque. 
 
  Strickly speaking, the total torque exerted on the planet should also include a contribution  from the cold thermal torque (Lega et al. 2014; Masset 2017).  The latter originates from dense lobes forming on each side of the orbit and is given by:

\begin{equation}
\Gamma_{\rm cold}=-1.61 \frac{\gamma-1}{\gamma}\frac{x_p}{\lambda_c}\Gamma_0
\end{equation}
where $x_p=\eta h^2 r_p$ is the distance to corotation, $\eta=\alpha/3+\beta/6+1/2$, and $\lambda_c=\sqrt{2\chi /3 \Omega_p \gamma}$ is the typical size of the lobes forming on either side of the orbit. 
The dashed blue line in Fig. \ref{fig:torques}  shows that cold thermal torque has a negligible effect on the location of the planet trap. It is interesting to note that inside the pressure bump,   $\eta<0$ such the cold thermal torque can become positive there. 

\subsubsection{Effect of taking into account dust torques}
\label{sec:dusttorques}

Previous work  has emphasized the ability for a low-mass planet to develop an asymmetric dust distribution in its vicinity, and which is responsible for a dust torque exerted on the planet (Benitez-Llambay et al. 2018). It has been shown that this dust torque can be potentially strong,  and  can even reverse inward planetary migration for cores with mass $\lesssim 10 M_\oplus$ (Guilera et al. 2023). Here, we examine how dust torques impact the planet trap concept mentioned in the previous section, and how they affect the trap location. We note in passing that the numerical setup adopted in this work leads to a resolution of $6\times 10^{-4}$ $r_p$ at the planet position,  similar to that used in the simulations of Benitez-Llambay et al. (2018) and Regaly (2020). It has been demonstrated by these authors  that such a resolution is sufficient to properly capture the dust dynamics in the vicinity of the planet (e. g. Regaly 2020). 

\begin{figure}
\centering
\includegraphics[width=\columnwidth]{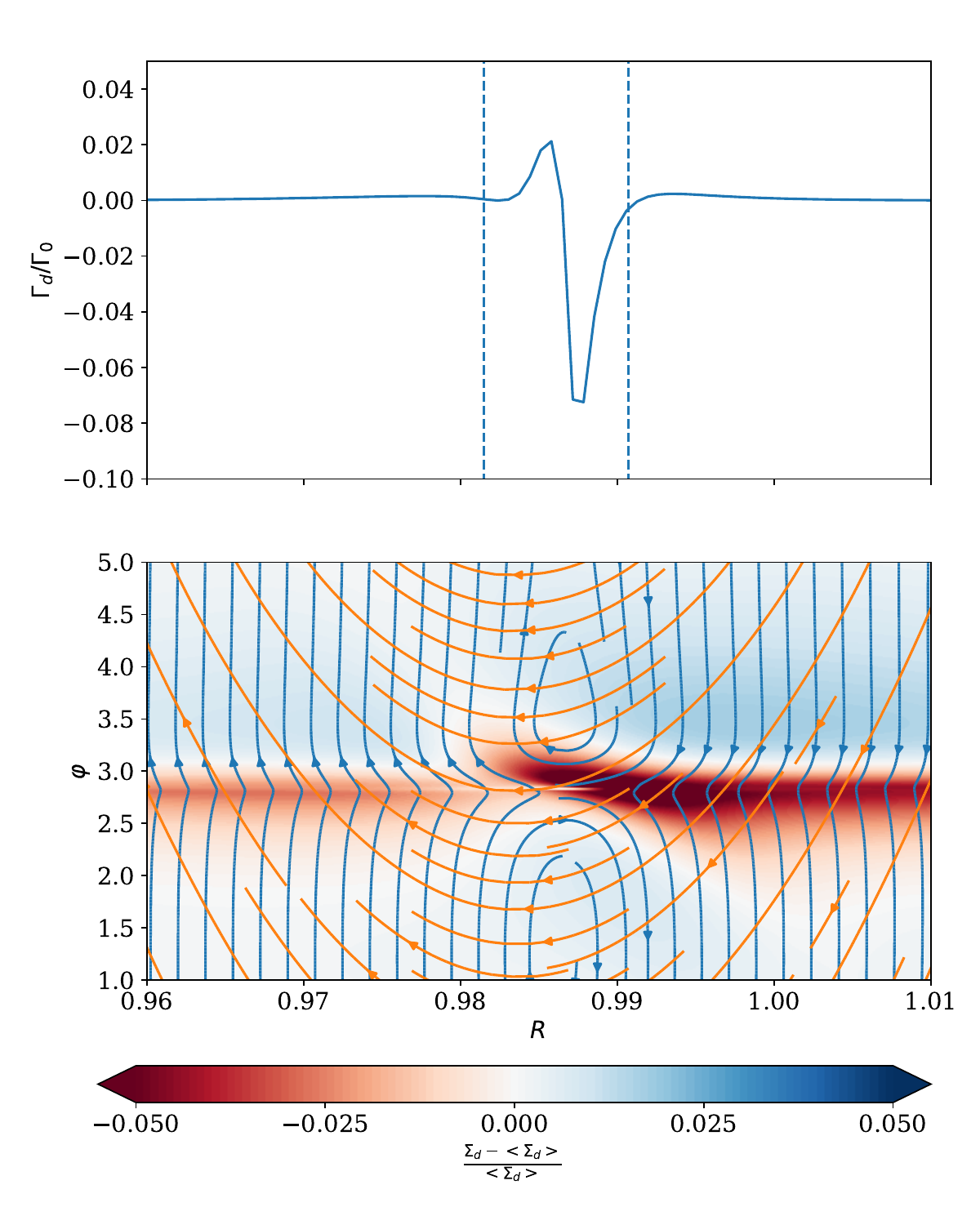}
\caption{Top panel: Pebble torque profile for $\st=0.01$ and $\alpha=10^{-4}$ in the situation where the planet can evolve under the effect of gas plus pebble torques. Bottom panel:  pebble surface density perturbation and streamlines (blue) for the same  run. Orange lines correspond to  streamlines obtained in absence of pressure bump. }
\label{fig:torque_peb}
\end{figure}

  Fig. \ref{fig:base}  shows that a planet trap is still effective when taking into account dust torques.The main difference is that the  location of the trap is slightly shifted inward, which suggests negative dust torques.  This is confirmed by examining the dust torque density which is shown as a function of radius in the upper panel of Fig. \ref{fig:torque_peb}.  Inspecting the time evolution of the dust torques also reveals that they tend to counterbalance  the positive gas torques at equilibrium.  For $\st=0.01$ and $m_p=0.1$ $M_\oplus$, a negative dust torque is in agreement with the results of Benitez-Llambay et al. (2018) (see their Fig. 2). However, whereas Benitez-Llambay et al. (2018) found that small planets feel a negative torque due to an overdensity behind the planet that competes with an overdensity located in front of the planet, here the origin of the negative torque exerted on the planet is found to be different. This is illustrated in the bottom panel of  Fig. \ref{fig:torque_peb} where we display the relative dust density perturbation near the planet. An overdense region behind the planet can indeed be observed but  a density depletion located in front of the planet also contributes to the negative dust torque acting on the planet. 
    
  In the bottom panel of  Fig. \ref{fig:torque_peb},  the overplotted blue streamlines also show possible horseshoe dynamics  in presence of a pressure bump, and which would give rise to a dust corotation torque acting on the planet. After executing their U-turn, pebbles migrating from the outer  to the inner disc  can drift back outside of the planet orbit as a result of the change in pressure gradient sign, leading to the formation of a corotation region. In absence of pressure bump, we see that  pebbles also execute U-turns as they approach the planet but the bended trajectories (orange lines in Fig. \ref{fig:torque_peb}) are not due to  horseshoe dynamics but occur simply because we employ a reference frame corotating with the embryo (Morbidelli \& Nesvorny 2012). 

By analogy with the classical corotation torque in a gas disc, we expect a dust corotation torque to be established provided that $\chi_\tau<1$ where $\chi_\tau$ is the ratio between the time $\tau_f$ for a pebble to move across the radial extent $x_s$ of the (dust) corotation region  due to the dust radial inflow and the libration timescale $\tau_{lib}$:
\begin{equation}
\chi_\tau=\frac{\tau_f}{\tau_{lib}}=\frac{3x_s^2\Omega}{8\pi a_p|v_{d,R}|}
\end{equation}
where $x_s$ is estimated to  $x_s\approx 5\times 10^{-3}a_p$ and $|v_{d,R}|\approx6\times 10^{-4}a_p\Omega$. This gives $\chi_\tau\approx 0.5$ such that we indeed expect the dust corotation torque to be   unsaturated. 

\subsubsection{Impact of pebble accretion}
\label{sec:pebble_accretion}
In this section, we study the impact of taking into account pebble accretion onto the core, which is modelled in our simulations using the procedure described in Sect. \ref{sec:accretion}.  As the planet accretes pebbles, release of potential energy heats the disc locally, leading to the formation of two underdense lobes  (Benitez-llambay \& Masset 2015)  that exert thermal  forces onto the planet.  Since our aim is to examine the effect of each physical process on the planet orbital evolution, torques resulting from pebble scattering and whose effect  has been discussed in the previous section are not included here. In addition,  the mass removed during the pebble accretion process is not added to the planet mass for the time being.

\begin{figure}
\centering
\includegraphics[width=\columnwidth]{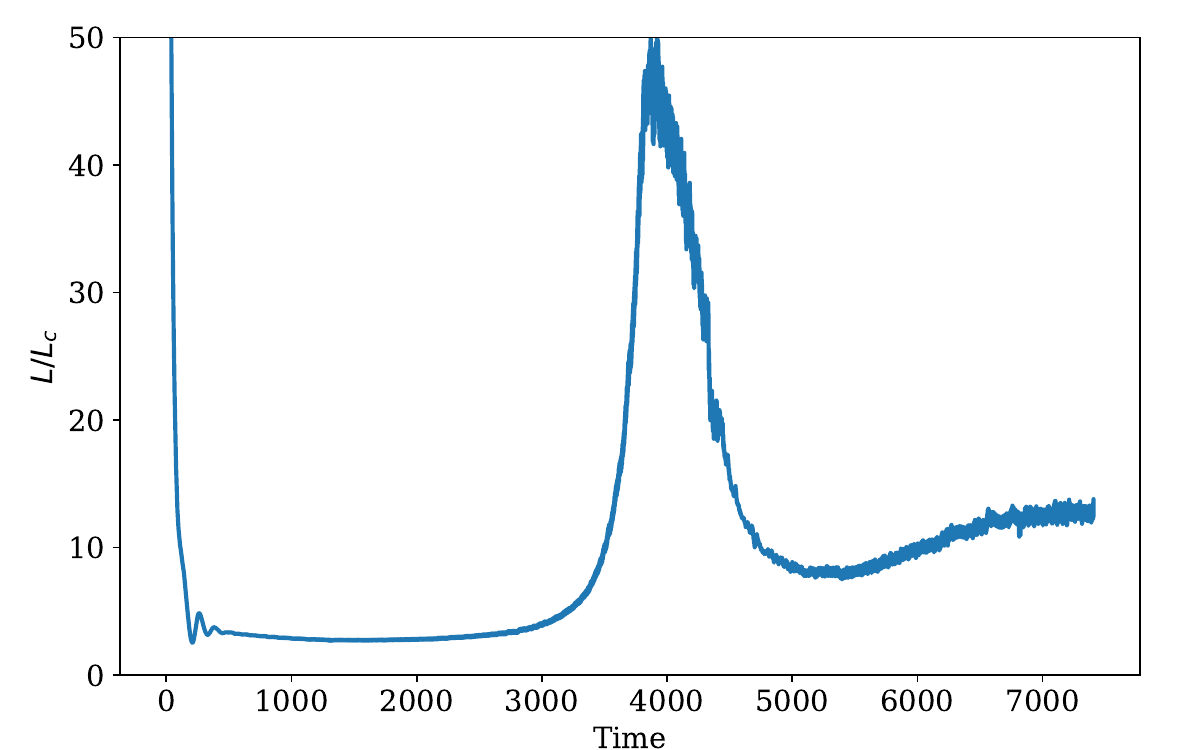}
\caption{Accretion luminosity as a function of time for the fiducial run  with $\st=0.01$, $\alpha=10^{-4}$ and pebble accretion included. For  luminosities higher than the critical value $L_c$, growth of planet eccentricity occurs as a result of thermal forces.}
\label{fig:lumin}
\end{figure}

 Under standard conditions, namely in discs where $\eta >0$, the torque resulting from accretion heat is positive. It has been shown that the net thermal torque, due to the contributions from the cold (negative) plus heating (positive) torques, is $>0$ provided that the planet luminosity exceeds a critical value $L_c$  given by (Masset 2017):
\begin{equation}
L_c=\frac{4\pi Gm_p \chi \rho}{\gamma}
\end{equation}
where $\chi$ is the thermal diffusion coefficient and $\rho$ the gas density. A planet with  super-critical luminosity tends to undergo not only ouward migration, but also eccentricity growth (Fromenteau \& Masset 2019) until the planet eccentricity saturates to a value close to the disc aspect ratio (Velasco Romero et al. 2022). \\
For a $0.1$ $M_\oplus$ planet forming at a pressure bump, the planet luminosity relative to $L_c$ is plotted as a function of time in Fig. \ref{fig:lumin}. Initially, $L/L_c\approx 3$ such that growth of eccentricity with time is expected. Referring back to Fig. \ref{fig:base}, this is indeed what is observed in the case with pebble accretion only (green line).  Linear theory predicts that this eccentricity  growth should be exponential as long as it is much smaller than the thermal lengthscale $\lambda_c$ (Masset 2017); 
whereas for a given value of the planet luminosity, the eccentricity tends to increase linearly with time once $e_p>> \lambda$ (Cornejo et al. 2023). This is referred to as the headwind-dominated regime where the two-lobes feature is replaced by  a hot trail (Chrenko et al. 2017). Here, $\lambda\approx 0.0015$ $a_p$ such that the episode of  rapid eccentricity growth that can be seen at Time $\approx 3\times 10^3$  is likely to arise in the headwind-dominated regime.  This is a consequence of two effects. First, it results from the feedback loop  between the eccentricity of the planet and its luminosity.  The pebble accretion efficiency increases with planet eccentricity while at the same time, the thermal forces that cause eccentricity growth strengthen as the planet becomes more luminous. Second, the $\eta$ parameter is continuously  increasing in absolute value as the planet migrates outward, resulting in a higher pebble flux. Although the semimajor axis evolution rate is much smaller than that of the eccentricity, the important increase in luminosity that is generated by both effects causes the brief period of fast outward migration that can also be observed at Time $\approx 3\times 10^3$.  As the eccentricity reaches $e_p\sim h$ at later times, the contribution from thermal forces to the semimajor axis evolution becomes negligible (Pierens 2023),  while both the resonant (Papaloizou \&  Larwood 2000) and corotation (Fendyke \& Nelson 2014) torques become quenched as well. This leads to a final evolution outcome with the  planet can evolve on  a fixed eccentric orbit. Here, the planet eccentricity at steady state is found to be $e_p\sim 0.04$ while $a_p\sim 1.025$, such that the planet is located just outside of the dust ring. 

\subsubsection{Combining pebble accretion and torques from pebble scattered flow}
\label{sec:acc+peb}

\begin{figure*}
\centering
\includegraphics[width=\textwidth]{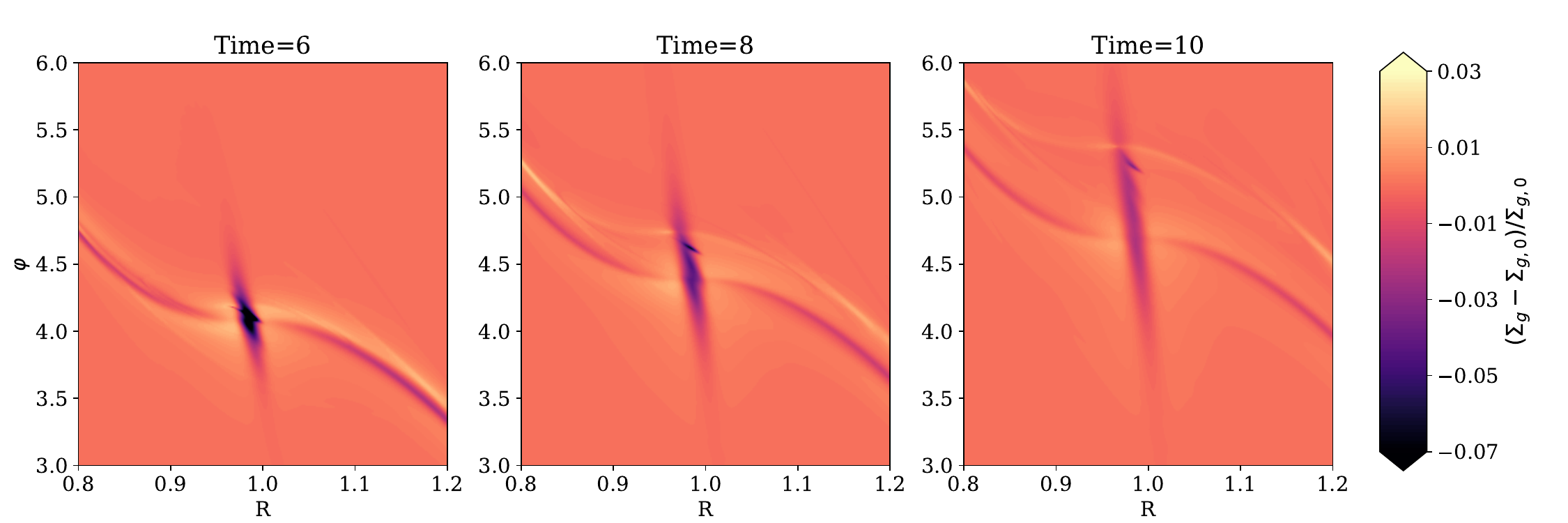}
\includegraphics[width=\textwidth]{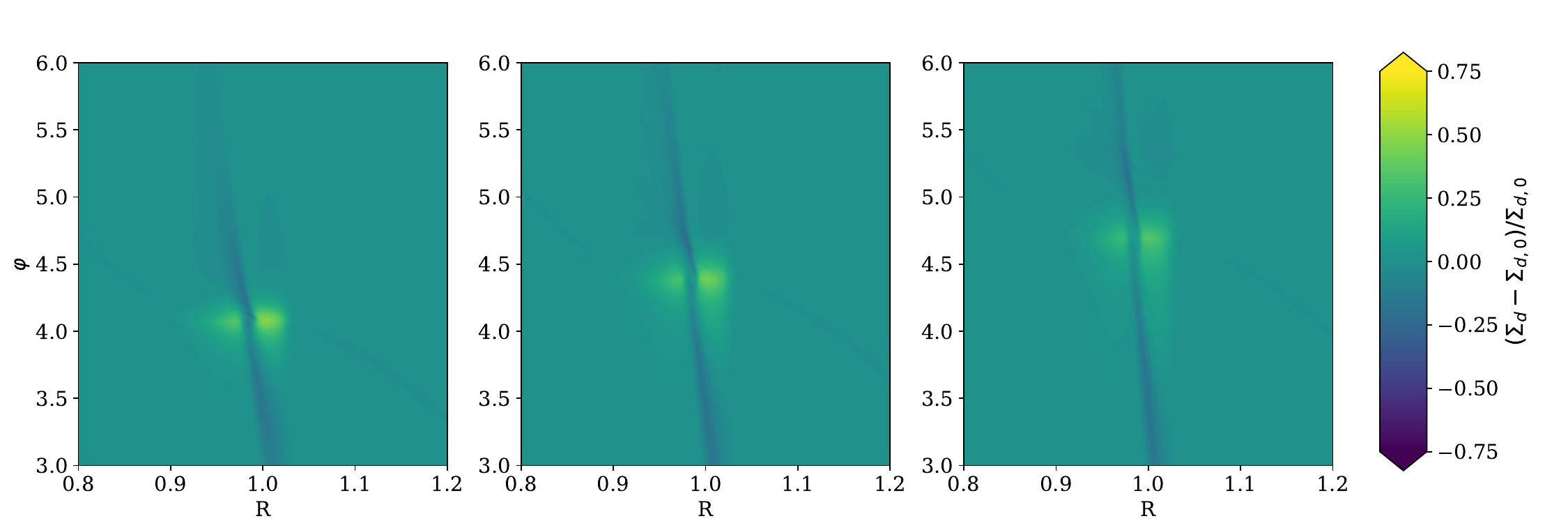}
\includegraphics[width=\textwidth]{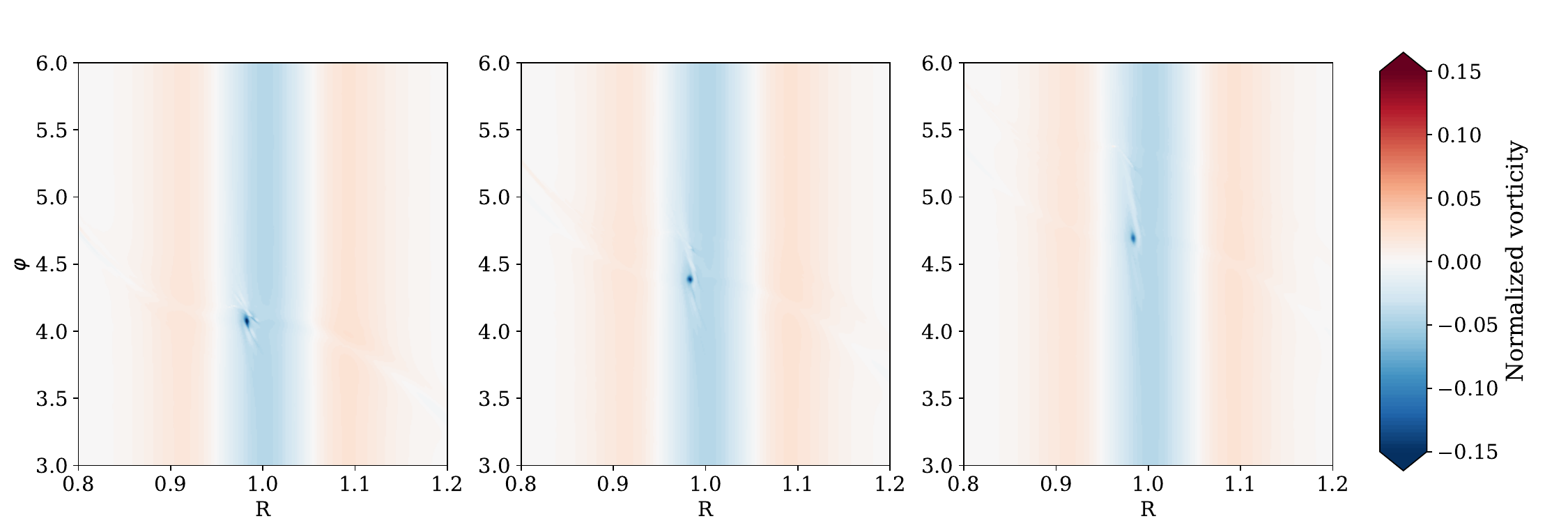}
\caption{Top panel: Gas surface density perturbation at the different times for the run with $\st=0.01$, $\alpha=10^{-4}$ and pebble accretion included. Gas heating due to pebble accretion leads to vortex formation at the planet position. Middle panel: Pebble surface density perturbation. Bottom panel: contours of the normalized vorticity for the same run.   }
\label{fig:rossby}
\end{figure*}

This situation is represented by the red line in Fig. \ref{fig:base} and resulted in a different mode of evolution. Here,  the planet is found to escape from the ring, migrating inward in the pebble-depleted region and maintaining a small eccentricity due to its small accretion luminosity. The continuous inward migration that is observed is clearly related to the significant increase in eccentricity that arises at early times. The large eccentricity that is  reached $e_p\sim 0.06$ quenches not only the corotation  (Fendyke \& Nelson 2014) but also the thermal  torques (Pierens 2023), which are both positive. In order to get insight into the physical origin of the eccentricity growth,  we show in the two upper panels of Fig.  \ref{fig:rossby} the gas and dust surface densities when the planet eccentricity begins to increase. The gas surface density reveals a vortex forming at the radial location of the planet, which is confirmed by inspecting the vortensity distribution in the third panel of Fig.  \ref{fig:rossby}. This is consistent with the recent results of Cummins et al. (2022) who found that including the thermal feedback onto the gas   resulting from the accretion luminosity tends to lead to the formation of an anticyclonic vortex. Consistently with Cummins et al. (2022), we also find that the vortex is formed {\it at} the location of the planet, as suggested by the Time=6 panel. It is important to note that vortex formation is also observed in the simulations corresponding to the previous situation where pebble accretion only was included, although it  had no impact on the planet orbital evolution because dust torques were not included in this run. This implies that the initial eccentricity growth that occurs is  due to interactions of the planet with the dust component of the vortex rather than with its gas component. We checked by inspecting the time evolution of the theoretical change of  $e_p$ due to the gas and dust components that this is indeed the case. Previous studies have also reported the plausibility  for a planet to undergo significant eccentricity growth due to its interaction with a dusty vortex (Pierens et al. 2019; Hsieh \& Lin 2020).  For instance Hsieh \& Lin (2020) found that the eccentricity  induced by a dusty vortex  is $e_p\sim 0.03$ in discs with metallicity $Z=0.3$
whereas $e_p\sim 0.1$ in disc with $Z\ge 0.5$. Since the planet is here located in a pressure with a dust-to-gas ratio $\approx 1$, a maximum eccentricity of $e_p\approx 0.06$ is therefore compatible with these estimations. 

\subsubsection{Effect of dust feedback}
\label{sec:feedback}
\begin{figure*}
\centering
\includegraphics[width=\textwidth]{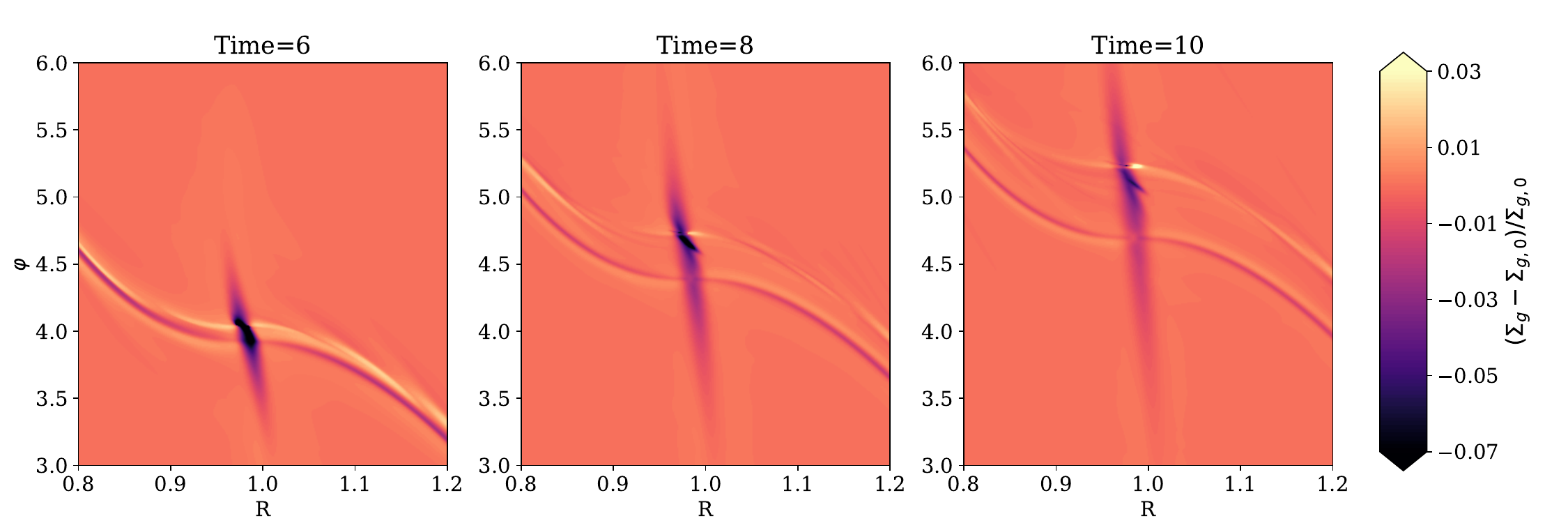}
\includegraphics[width=\textwidth]{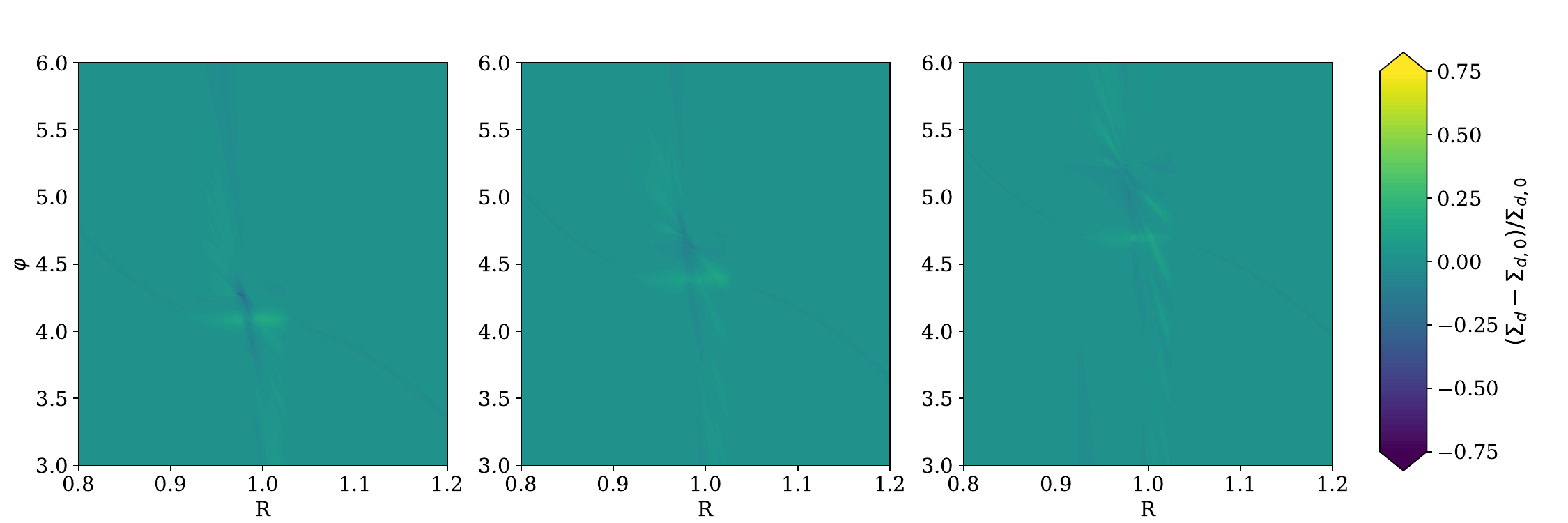}
\includegraphics[width=\textwidth]{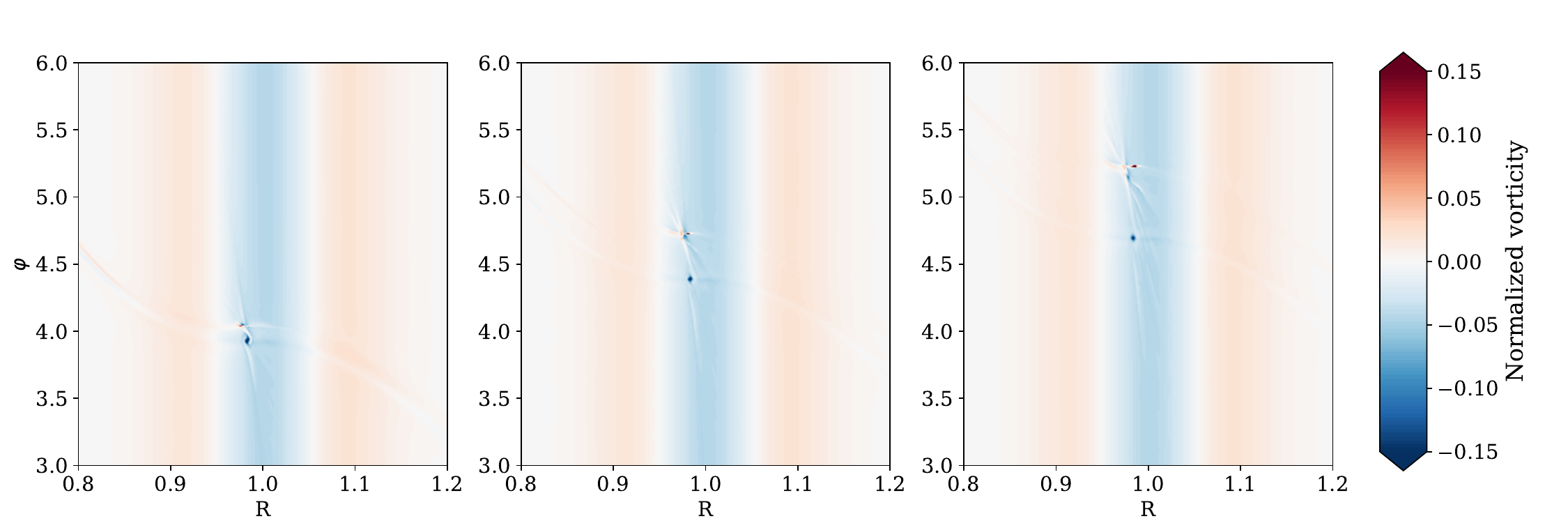}
\caption{Same as Fig. \ref{fig:rossby} but with the effect of dust back-reaction onto the gas included.}
\label{fig:rossby_feedback}
\end{figure*}

Due to the large solid abundance within the ring, an important issue that needs to be examined  is the effect of dust feedback onto the gas. Compared to the previous case where it was discarded, including dust feedback (purple line in Fig. \ref{fig:base}) enables the planet to remain trapped near the pressure maximum, at a location similar to that found in the situation where only torques due to pebble scattering were considered (orange line in Fig. \ref{fig:base}). As first glance this is surprising since it has been shown that by altering the gas azimuthal velocity, dust feedback tends to lead to meso-scale instabilities (Pierens et al. 2019; Huang et al. 2020; Hsieh \& Lin 2020), which can subsequently excite the planet's eccentricity.  This is especially true in the case where vortices are formed as a result of the Rossby Wave Instability (RWI) associated with an extremum in the potential vorticity. Here, however, the source of vorticity is a baroclinic instability associated with the embryo accretion luminosity (Owen \& Kollmeier 2017) rather than the RWI. In this case,  it is plausible that vortices suffer drag instabilities while they collect dust, as demonstrated by past studies (Fu et al. 2014; Crnkovic-Rubsamen et al. 2015; Surville \& Mayer 2019). Our results are consistent with this scenario, since we find that including the effect of dust back-reaction leads to weaker vortices. This can be clearly observed by comparing Figs. \ref{fig:rossby} and \ref{fig:rossby_feedback} where we show the density and vorticity distributions when including dust feedback. An additional difference is that the amplitude of the hot trail is higher when dust feedback is included. Given that it is responsible for driving its eccentricity, this causes the planet to reach a maximum value $e_p\sim h$ much earlier in comparison with the situation discussed in Sect.  \ref{sec:pebble_accretion} where the effect of the dust backreaction onto the gas was not included. The effect of the thermal torque on the semi-major axis also weakens more rapidly,  with the consequence that the main contribution to the dust torque becomes dust scattering. At equilibrium, the planet eccentricity is therefore driven by thermal forces whereas the semi-major axis evolution is driven by dust scattering.   

\subsection{Taking into account the change in planet mass due to pebble accretion}

\begin{figure}
\centering
\includegraphics[width=\columnwidth]{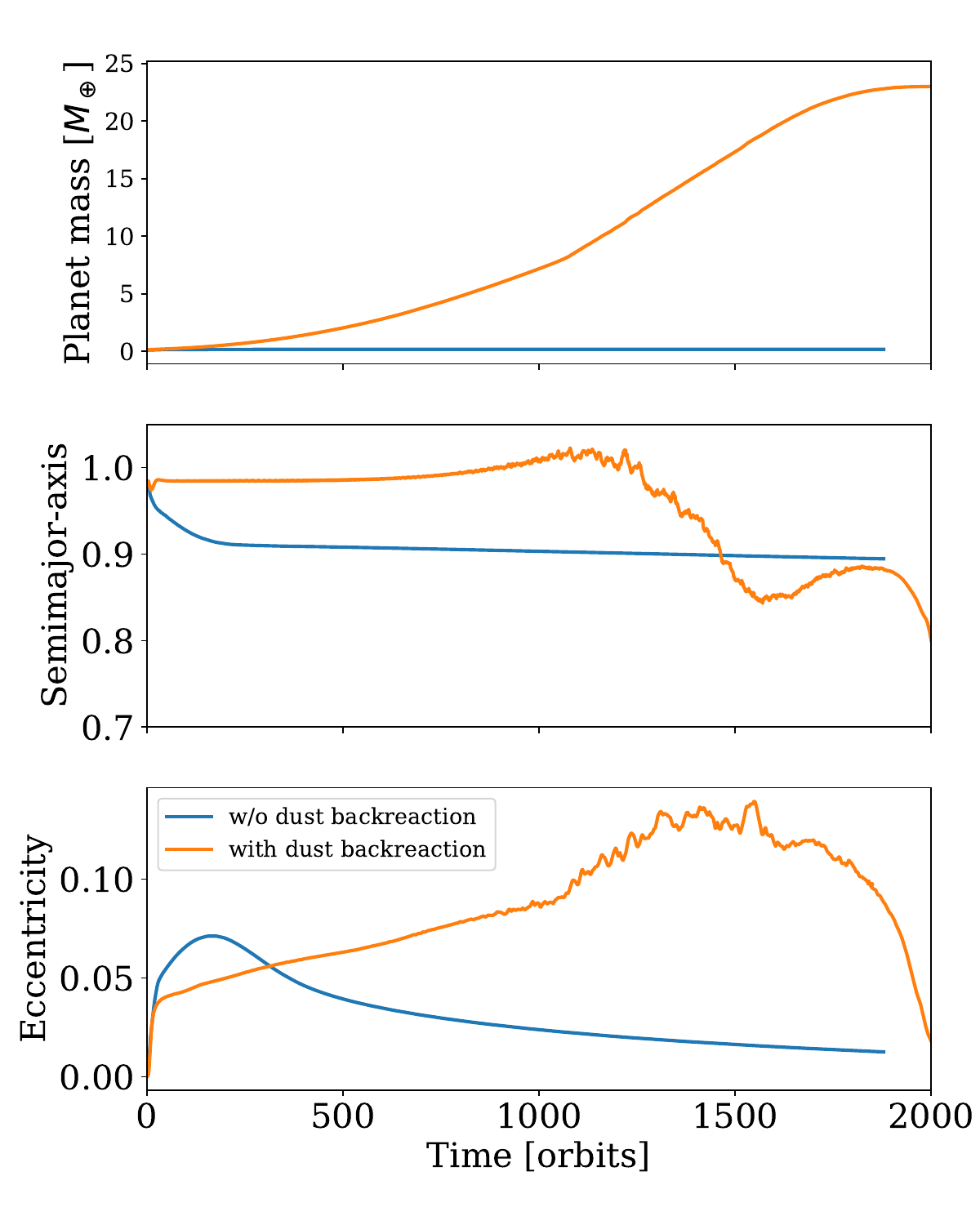}
\caption{Planet mass, semi-major axis and eccentricity as a function of time for runs in which the planet grows as a result of pebble accretion.}
\label{fig:growth}
\end{figure}

\begin{figure}
\centering
\includegraphics[width=\columnwidth]{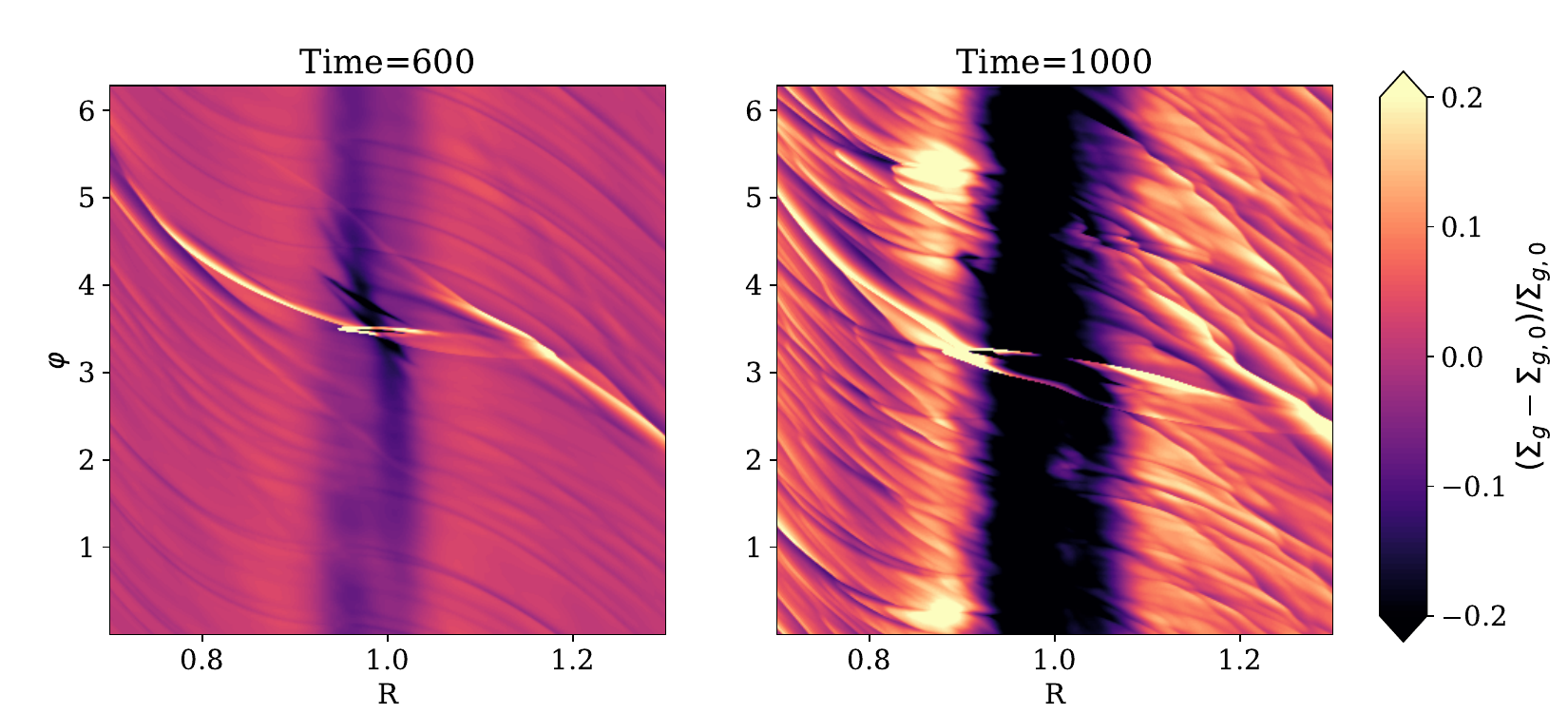}
\includegraphics[width=\columnwidth]{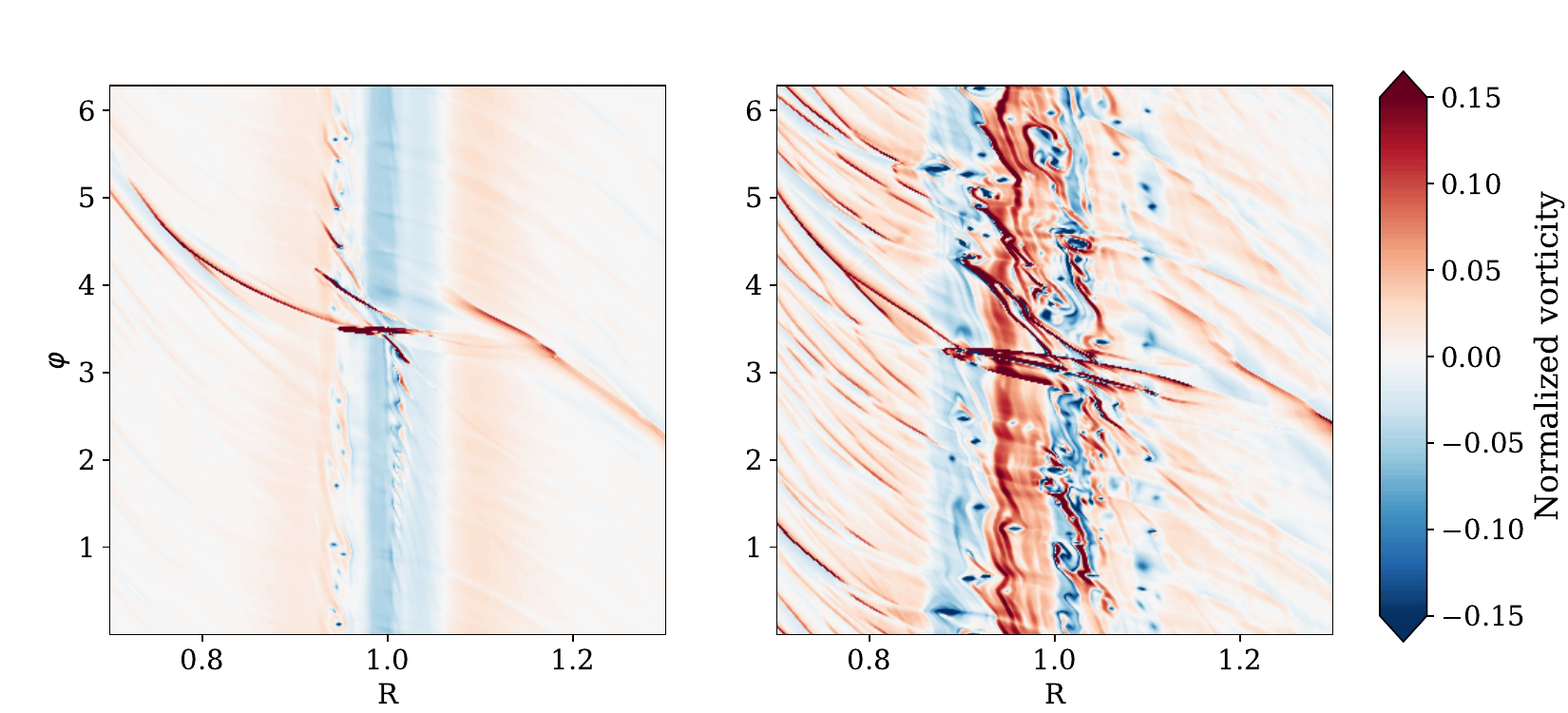}
\caption{Top panel: Gas surface density perturbation at the different times for the run where the planet is allowed to grow and with the effect of dust feedback onto the gas is included. Bottom panel: contours of the normalized vorticity at the same times.  }
\label{fig:vortex}
\end{figure}

We now discuss the results of more realistic simulations in which planets are truly allowed to accrete pebbles and increase in mass.  In that case, Fig. \ref{fig:growth} reveals the important role played by the dust backreaction on  the orbital and mass evolution of the planet. When this effect is discarded, the evolution is  similar to that described in Sect. \ref{sec:acc+peb}, involving rapid escape from the pressure bump due vortex formation at the planet position. Because pebbles accumulate at the pressure maximum, planet growth is mostly  stalled as it migrates inward in such a way that the maximum mass of the growing embyo is only $\approx 0.2$ $M_\oplus$. When the effect of dust feedback onto the gas is included, however, the growth of the planet is also found to stop, but to a much higher mass of $\approx 20$ $M_\oplus$. This demonstrates that the growth of giant planet cores at pressure bumps is plausible, and can happen on a relatively short timescale. Again, the early evolution in that case proceeds similarly to what has been described in Sect. \ref{sec:feedback}, with the planet trapped at the zero-torque radius set by gas torques and pebble scattering. Because this location is close to the pressure maximum, this allows the planet to accrete at a higher rate and to rapidly reach a mass where gap opening can occur. Not surprisingly, the gap opening process is accompanied  by vortex formation, as can be seen in Fig. \ref{fig:vortex} where we show the gas density and vorticity at two different times.  At Time=$1000$, several vortices form at  the inner edge of the gap carved by the planet at $R\sim 0.9$. Given that the planet has semimajor axis $a_p\approx 1$ and eccentricity $e_p\sim 0.1$ at that time, this implies significant planet-vortex interactions when the planet is at pericentre. This cause the planet to subsequently undergo an episod of rapid inward migration (Lin \& Papaloizou 2010; Wafflard-Fernandez \& Baruteau 2020) and escape from the pressure bump. From this point onward, pebble accretion is stopped due to i) the very low density of pebbles in the inner disc while the planet continuously migrate inward and ii) the halting of pebble flux at the gap created by the planet. 

\begin{figure*}
\centering
\includegraphics[width=0.45\textwidth]{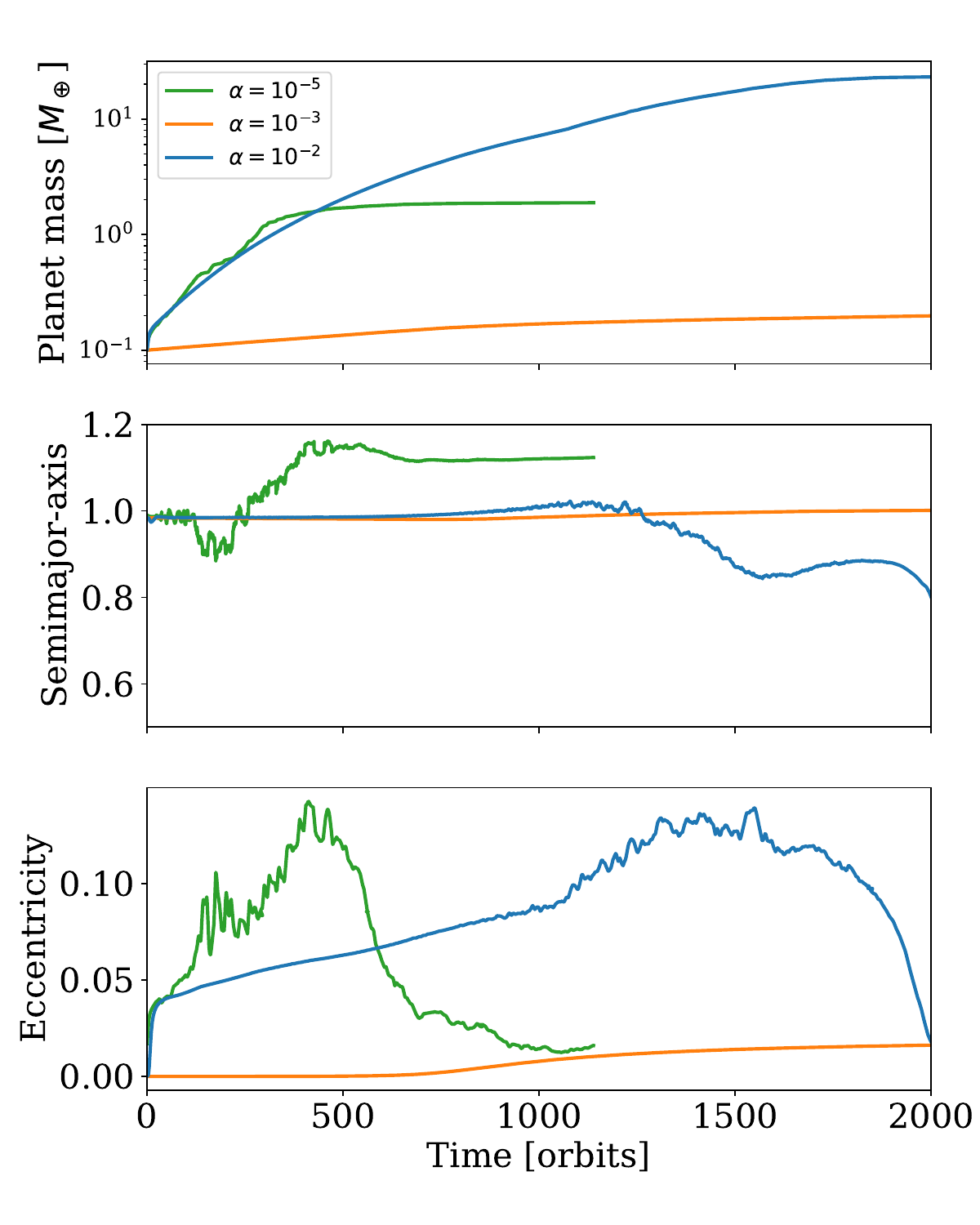}
\includegraphics[width=0.45\textwidth]{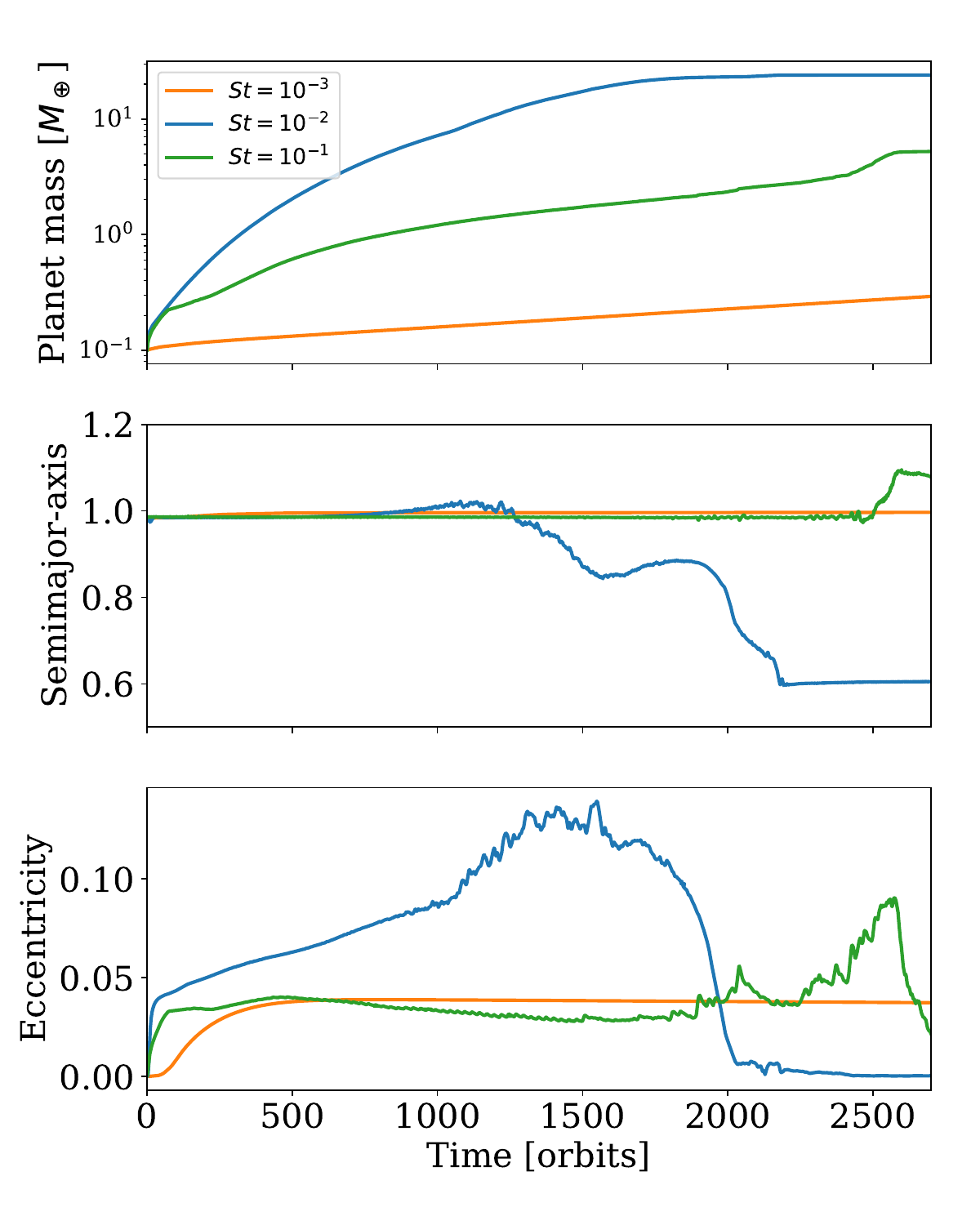}
\caption{Left: planet mass, semi-major axis and eccentricity as a function of time for runs in which the planet grows as a result of pebble accretion and for different values of the $\alpha$ viscosity and for $\st=0.01$. Right: same but for different values of the the Stokes number and for $\alpha=10^{-4}$.  }
\label{fig:vary_alpha}
\end{figure*}

\subsection{Varying the $\alpha$ viscosity}
In the left panel of Fig. \ref{fig:vary_alpha}, we present the time evolution of planet mass, semimajor axis, and eccentricity for three different values of the $\alpha$ viscosity. 
Here, dust torques have been taken into account, as well as the effect of dust feedback onto the gas. Clearly, the value of the viscosity  has a major impact on the planetary growth process. For $\alpha=10^{-3}$, pebble accretion is slow at first because of the larger width of the pressure bump (see Fig. \ref{fig:profiles_alpha}), resulting in weaker dust trapping. This also occurs because gas diffusivity tends to hinder eccentricity growth, especially for smaller planetary masses (Chameta et al. 2021). As pebble accretion efficiency scales with planet eccentricity, pebble accretion is therefore expected to be less efficient  for high values of the gas diffusivity.  \\
One might expect that the planet would grow more rapidly in the run with $\alpha=10^{-5}$. However, this is not observed in our simulations, as the planet mass saturates to $m_p\approx 0.5$ $M_\oplus$, whereas it reaches $m_p\approx 20$ $M_\oplus$ when $\alpha=10^{-2}$. Although not shown here, we find that this is due to the vortices forming in the vicinity of the planet.   These vortices generate spiral waves that propagate outward and dissipate. For $\alpha=10^{-5}$, this leads to the formation of a secondary pressure bump that can trap pebbles drifting from the outer disc, with the consequence that the planetary growth process is stopped. 

\subsection{Varying the Stokes number}

\begin{figure}
\centering
\includegraphics[width=\columnwidth]{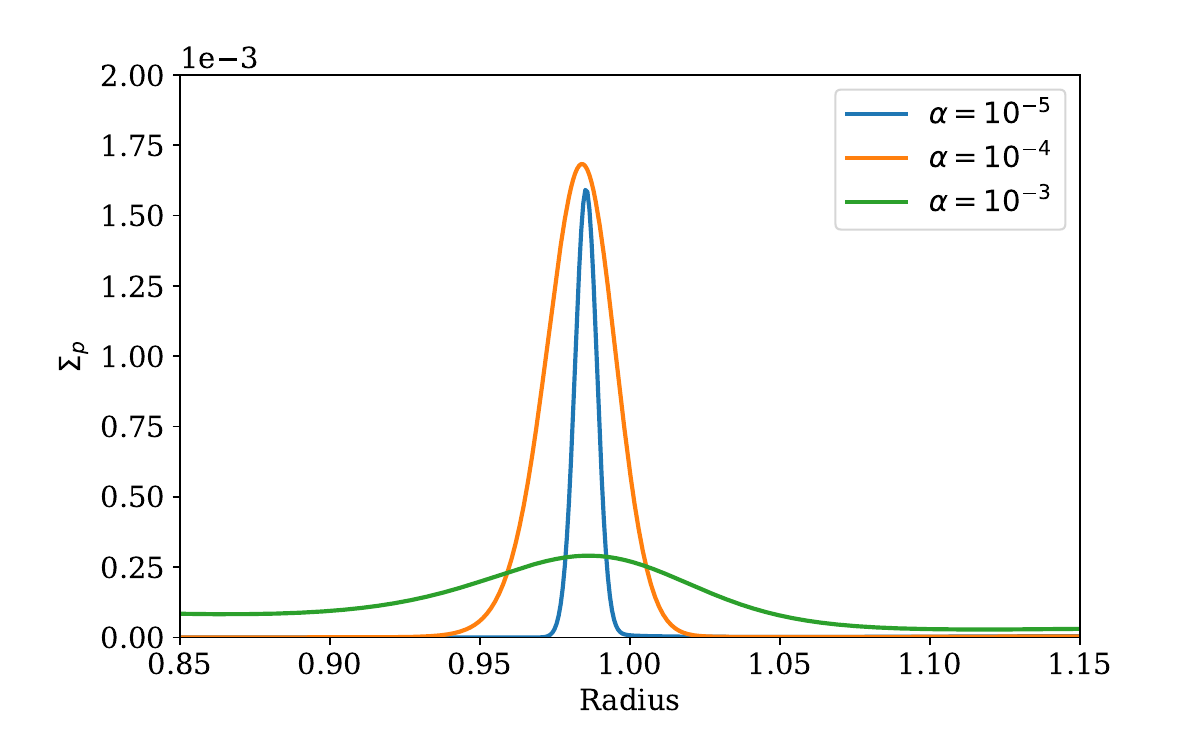}
\includegraphics[width=\columnwidth]{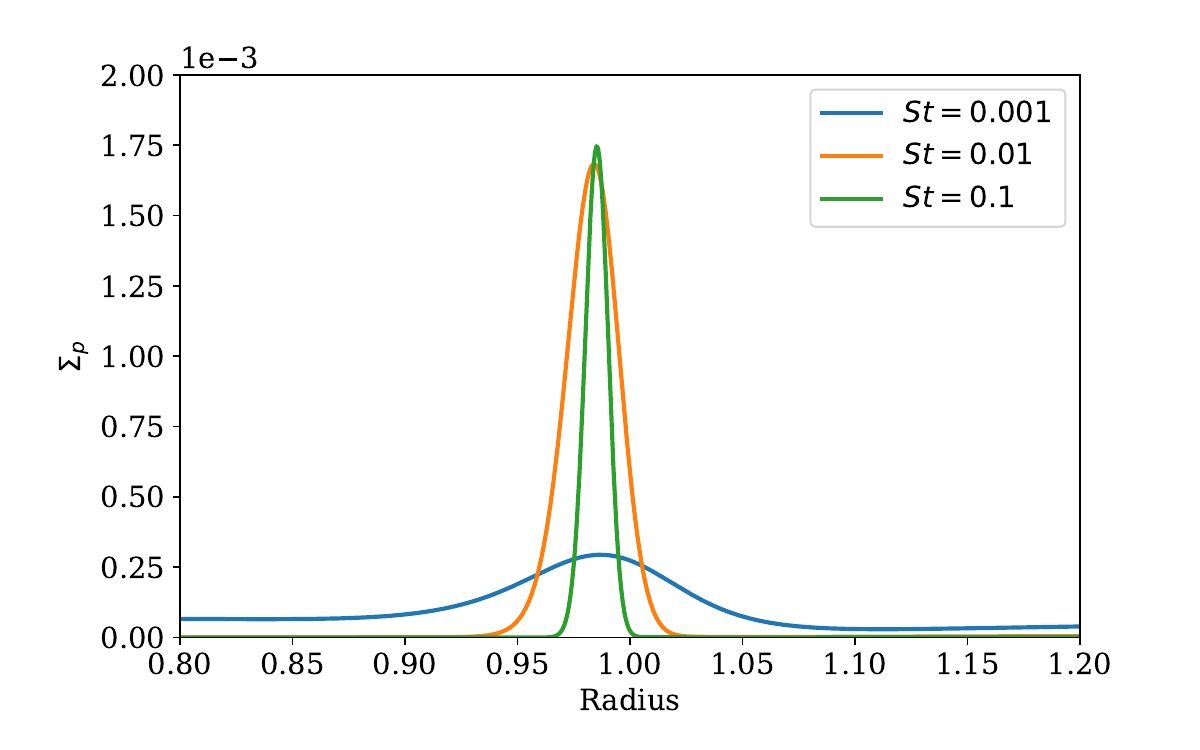}
\caption{Top: pebble surface density $\Sigma_p$ profile for different values of the $\alpha$ viscosity. Bottom: pebble surface density $\Sigma_p$ profile for different values of the Stokes number $\st$. }
\label{fig:profiles_alpha}
\end{figure}

\begin{figure}
\centering
\includegraphics[width=0.45\textwidth]{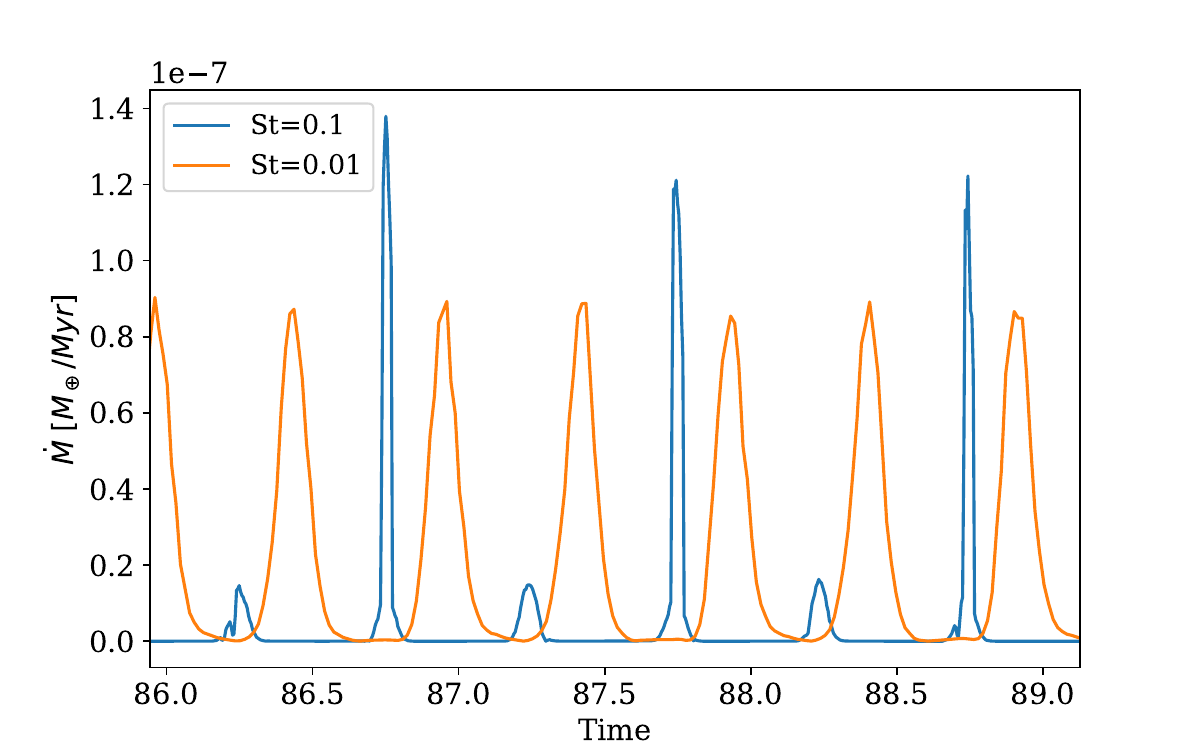}
\caption{Accretion rate onto the planet over a few orbits for the runs with $\st=0.1$ and $\st=0.01$. }
\label{fig:mdot}
\end{figure}

We now turn our attention to the dependence of our results on the Stokes number.  The pressure bump profiles at quasi steady-state are shown, for different values of $\st$, in the bottom panel of Fig. \ref{fig:profiles_alpha}; whereas the corresponding time evolution of planet mass, semimajor axis and eccentricity are presented in the right panel of Fig.   \ref{fig:vary_alpha}. Compared to the case with $\st=0.01$, the mass growth rate is  significantly reduced for $\st=0.001$. Here, this is simply a consequence of the less pronounced density peak for this value of the Stokes number.  In absence of a pressure bump,   we would indeed have  expected a higher accretion rate, as $\st=0.001$ pebbles are more coupled to the gas and drift more slowly, such that they have a higher probability of being accreted by the planet (Velasco-Romero et al. 2022). \\
The smaller pebble accretion rate that found for  $\st=0.1$ (relative to  $\st=0.01$) is  due to both larger drift velocity, and the  narrower  dust ring. Provided that the width of the dust ring is small enough,   we indeed find that the planet can eventually evolve outside of the dust ring over one orbital period due to  its  finite eccentricity. To demonstrate that this is indeed the case, we show in Fig. \ref{fig:mdot}  the accretion rate onto the planet over a few orbits. For $\st=0.01$, the accretion rate oscillates with a period corresponding to half of the planet orbital period whereas for $\st=0.1$,  the oscillation period corresponds to that of the planet.  In particular, the  accretion rate is maximum when the planet is at pericentre, whereas it takes small values at apocentre. This corresponds to a situation where it temporarily escapes from the dust ring. For $\st=0.01$, such large variations in the pebble accretion rate are not observed, simply because the dust ring is wider in that case. \\
Similarly to the case $\st=0.01$, the planet mass is also found to saturate for $\st=0.1$, but at a somewhat smaller value $m_p\approx 5$ $M_\oplus$.  Here, pebble accretion is stopped once the planet mass becomes high enough to significantly modify the pebble disc structure. From this time, the amplitude of the density peak is smaller which, combined with the significant value for the planet eccentricity $e_p\sim 0.05-0.1$ leads to a drastic reduction in the pebble accretion rate.  This is consistent with the results of Liu \& Ormel (2018) who found for similar parameters, $e_p\sim 0.05-0.1$ and $\st=0.1$, a very small pebble accretion efficiency (see their Fig. 5).

\section{Conclusions}

In this paper, we studied the  orbital evolution of low-mass planets embedded in pressure bumps.  Due to their ability to trap drifting pebbles, pressure bumps have been considered as favoured locations for planetesimal formation through the streaming instability.  We performed 2-dimensional, two-fluid hydrodynamical simulations in which a  Gaussian pressure bump is modelled by adjusting the viscosity profile of the disc. The planet has an initial mass of $0.1$ $M_\oplus$ and can grow by accreting pebbles from the dust ring, with a pebble accretion efficiency that is calculated using the prescription of Liu \& Ormel (2018). Heat release resulting from pebble accretion is taken into account, such that planet is subject to  thermal torques. We also take into account the feedback between eccentricity and luminosity, as well as the torques resulting from the pebble scattered flow (Benitez-Lambay et al. 2018). 

We found evidence for the existence of a dust corotation torque,  resulting from the change in the dust streamlines at the pressure bump. This additional corotation torque can slightly alter the location of the planet trap, in comparison with the situation where gas torques only are considered. We also found that due to the significant pebble surface density at the  bump location, the accretion luminosity of a $0.1$ $M_\oplus$ protoplanet can easily exceed the critical luminosity above which eccentricity growth is expected as a result of thermal forces. Eccentricity growth causes both Lindblad and corotation torques to weaken,  and the planet ultimately end up moving just outside of the dust ring, on an eccentric orbit with $e_p\approx h$, where $h$ is the disc aspect ratio. Thermal feedback onto gas also leads to the formation of an anticyclonic vortex at the planet location, consistently with the results of Cummins et al. (2022). This vortex can subsequently capture  drifting pebbles and we find that the interaction with the dusty vortex can make the planet escape from the pressure bump when  dust back-reaction is not accounted for. Including the effect of dust feedback onto the gas, however,  weaken the dusty vortex, which enable the planet to achieve a stable, eccentric orbit close to the gas pressure maximum.

Simulations in which the planet mass is allowed to increase as a consequence of pebble accretion resulted in the formation of giant planet cores with mass in the range $5-20$  $M_\oplus$, depending on the values for the Stokes number $\st$ and $\alpha$ viscosity. Pebble accretion was particularly efficient at the pressure bump for  $\st=0.01$ and $\alpha = 10^{-4}$. For smaller values of $\st$ the mass of the dust ring is simply too small to enable strong pebble accretion whereas for  higher $\st$, pebble accretion efficiency decreases due to a larger drift velocity. Moreover, we find in that case that the  planet can temporarily  evolve outside of  the dust ring as it approaches its pericentre, resulting in a reduced accretion rate.  

 In protoplanetary discs with  $\alpha \gtrsim 10^{-4}$, the planet luminosity tends to decrease below the critical luminosity (Velasco-Romero et al. 2022),  thereby preventing eccentricity growth through thermal forces and consequently efficient pebble accretion. Similarly, for $\alpha \lesssim 10^{-5}$, the planet achieves a maximum mass of  just $\approx 0.5$ $M_\oplus$, due to the formation of strong vortices which tend to create a secondary pressure bump outside the planet orbit and where the pebble flux is halted. 
 
 One main limitation here is that we only performed 2D simulations and future work should adopt a more realistic 3D setup. It has been indeed shown that the streamlines outflowing from the Hill sphere may be significantly distorted in 3D (Chrenko \& Lambrechts 2019) such that we expect significant differences in the heating torque magnitude between 2D and 3D.
\section*{Acknowledgments}
Computer time for this study was provided by the computing facilities MCIA (M\'esocentre de Calcul Intensif Aquitain) of the Universite de Bordeaux and by HPC resources of Cines under the allocation A0150406957 made by GENCI (Grand Equipement National de Calcul Intensif).  SNR acknowledges funding from the French Programme National de Plan\'etologie (PNP), and in the framework of the Investments for the Future programme IdEx, Universit\'e de Bordeaux/RRI ORIGINS. 
\section*{Data Availability}

The data underlying this article will be shared on reasonable request to the corresponding author.

\begin{appendix}

\section{Pebble accretion efficiency for eccentric planets}

In this section, we give the prescriptions of Liu \& Ormel (2018) and Ormel \& Liu (2018) for the pebble accretion efficiency of eccentric planets and that we used in our simulations. The pebble accretion efficiency on the 2D settling regime is given by:

\begin{equation}
\epsilon_{2D,\rm set}=0.32\sqrt{\frac{q}{\st \;\eta^2}\frac{\Delta V}{v_k}}f_{\rm set}
\label{eq:a1}
\end{equation}
where $v_k$ is the keplerian velocity and $\Delta V$  the pebble-embryo relative velocity which is given by:

\begin{equation}
\Delta V=\max(V_{\rm circ}, V_{\rm ecc})
\end{equation}

with:

\begin{equation}
V_{\rm ecc}=0.76e_p v_k
\end{equation}

and:

\begin{equation}
V_{\rm circ}=\left[1+5.7\left(\frac{q\st}{\eta^3}\right)\right]^{-1}+0.52(q\st)^{1/3}v_k
\end{equation}

Moreover, in Eq. \ref{eq:a1}, the transition function is given by:
\begin{equation}
f_{\rm set}=\exp\left[-0.5\frac{\Delta V}{V_\ast^2}\right]\times\frac{V_\ast}{\sqrt{V_\ast^2+0.33\sigma_{pz}^2}}
\end{equation}

where $V_\ast$ is the transition velocity:
\begin{equation}
V_\ast=\left(\frac{q}{\st}\right)^{1/3}v_h
\end{equation}

and $\sigma_{pz}$ is the vertical turbulent velocity (Youdin \& Lithwick 2007):

\begin{equation}
\sigma_{pz}=\frac{\alpha}{1+\st}\left(1+\frac{\st}{1+\st}\right)^{-1/2}hv_k
\end{equation}

The pebble accretion in the 3D settling regime is given by:

\begin{equation}
\epsilon_{3D,\rm set}=0.39\frac{q}{\eta h_d}f_{\rm set}^2
\end{equation}

with $h_d$ the pebble disc aspect ratio  (Youdin \& Lithwick 2007):

\begin{equation}
h_d=\sqrt{\frac{\alpha}{\alpha+\st}}\left(1+\frac{\st}{1+\st}\right)^{-1/2}h
\end{equation}

 In this regime, the accretion radius $R_{acc}$  is given by:

\begin{equation}
R_{acc}=\sqrt{\frac{Gm_p t_s}{\Delta V}},
\end{equation}

Finally, the accretion efficiency in the 2D and 3D ballistic regimes are respectively given by:
\begin{equation}
\epsilon_{2D,\rm bal}=\frac{R_p}{2\pi \eta R \st}\sqrt{\frac{2qR}{R_p}\left(\frac{\Delta V}{v_k}\right)^2}(1-f_{\rm set})
\end{equation}

and:
\begin{equation}
\epsilon_{3D,\rm bal}=\frac{1}{4\sqrt{2\pi} \eta h_d \st}\left(2d\frac{v_k}{\Delta V}\frac{R_p}{R}+\frac{R_p^2}{R^2}\frac{\Delta V}{v_k}\right)(1-f_{\rm set}^2)
\end{equation}

In the ballistic regime, $R_{acc}$ reads:

\begin{equation}
R_{acc}=R_p\sqrt{\left(\frac{V_{esc}}{\Delta V}\right)^2+1}
\end{equation}

where $V_{esc}=\sqrt{\frac{2Gm_p}{R_p}}$ is the escape velocity.\\

The expression that we employ to calculate the pebble accretion efficiency in our calculations is given by:

\begin{equation}
\epsilon=\frac{f_{\rm set}}{\sqrt{\epsilon_{2D,\rm set}^{-2}+\epsilon_{3D,\rm set}^{-2}}}+\frac{1-f_{\rm set}}{\sqrt{\epsilon_{2D,\rm bal}^{-2}+\epsilon_{3D,\rm bal}^{-2}}}
\end{equation}
\end{appendix}

\end{document}